%% file: main.tex
\pdfoutput=1
\newif\iffull\fulltrue

\documentclass[conference]{IEEEtran}
\IEEEoverridecommandlockouts
\usepackage{cite}
\usepackage{amsmath,amssymb,amsfonts}
\usepackage{algorithmic}
\usepackage{graphicx}
\usepackage{textcomp}
\usepackage{xcolor}
\def\BibTeX{{\rm B\kern-.05em{\sc i\kern-.025em b}\kern-.08em
    T\kern-.1667em\lower.7ex\hbox{E}\kern-.125emX}}

 \makeatletter %
 \renewcommand\@oddhead{\thepage}
\usepackage{style}
\input{macro}

\begin{document}
\bstctlcite{IEEEexample:BSTcontrol}
\title{Parallel Filtered Graphs for Hierarchical Clustering\\
}

\author{\IEEEauthorblockN{Shangdi Yu}
\IEEEauthorblockA{
\textit{MIT CSAIL}}
\and
\IEEEauthorblockN{Julian Shun}
\IEEEauthorblockA{
\textit{MIT CSAIL}
}
}
\IEEEaftertitletext{\vspace{-1.2\baselineskip}}

\maketitle

\input{abstract.tex}

\input{intro}

\input{fig_example}

\input{background}
\input{related}
\input{tmfg}
\input{dbht}

\input{analysis}

\input{exp}

\input{conclusion}

\section*{Acknowledgements}
This research is supported by
DOE Early Career Award \#DE-SC0018947,
NSF CAREER Award \#CCF-1845763, Google Faculty Research Award, Google Research Scholar Award, 
FinTech@CSAIL Initiative, DARPA
SDH Award \#HR0011-18-3-0007, and Applications Driving Architectures
(ADA) Research Center, a JUMP Center co-sponsored by SRC and DARPA.

\bibliographystyle{IEEEtran}

\iffull
\newpage

\appendix\label{sec:appendix}

\begin{figure}[t]
    \centering
    \includegraphics[width = \columnwidth]{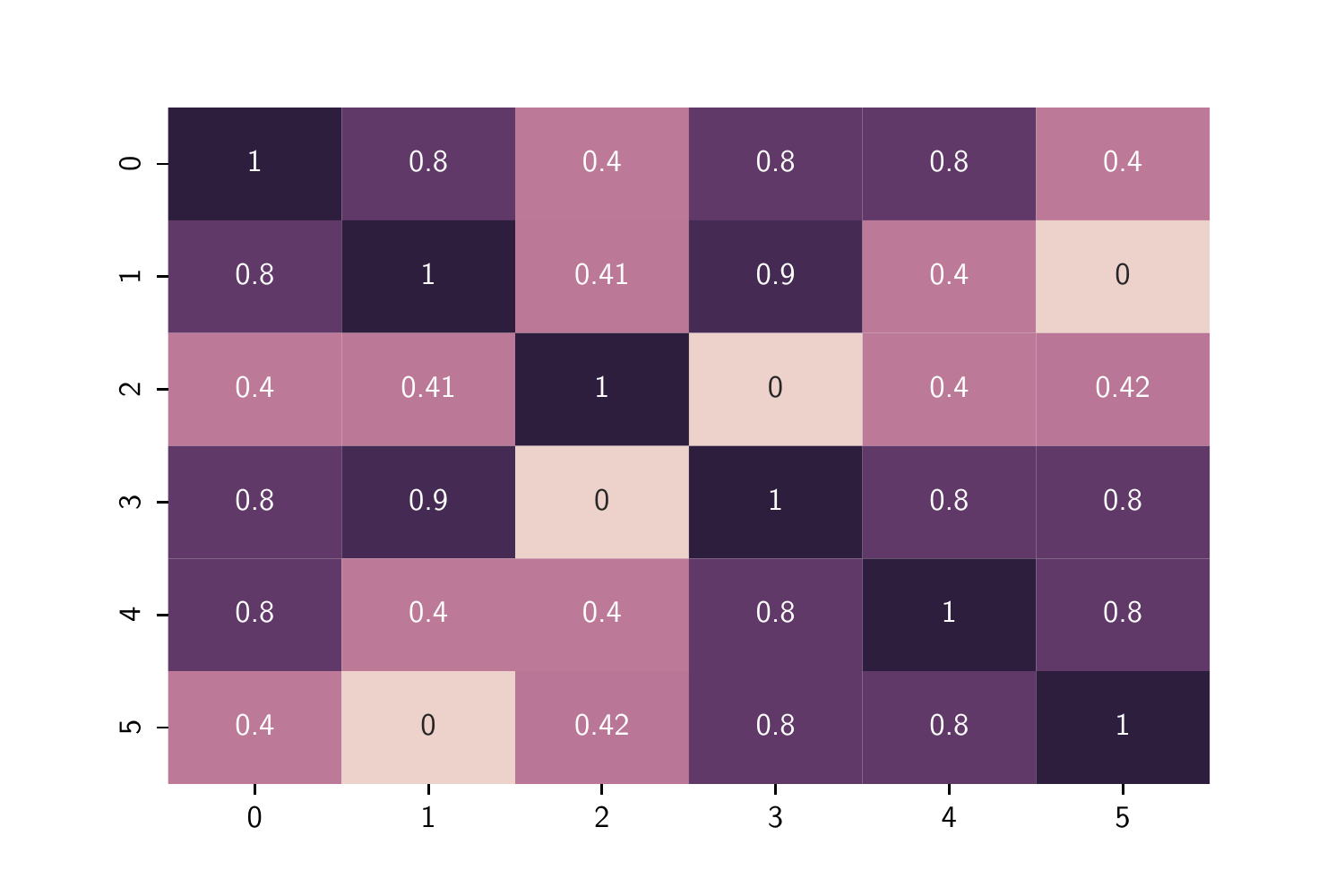}
    \caption{Correlation matrix between data points.}
    \label{fig:heatmap}
\end{figure}

\begin{figure}[t]
    \centering
    \includegraphics[width = \columnwidth]{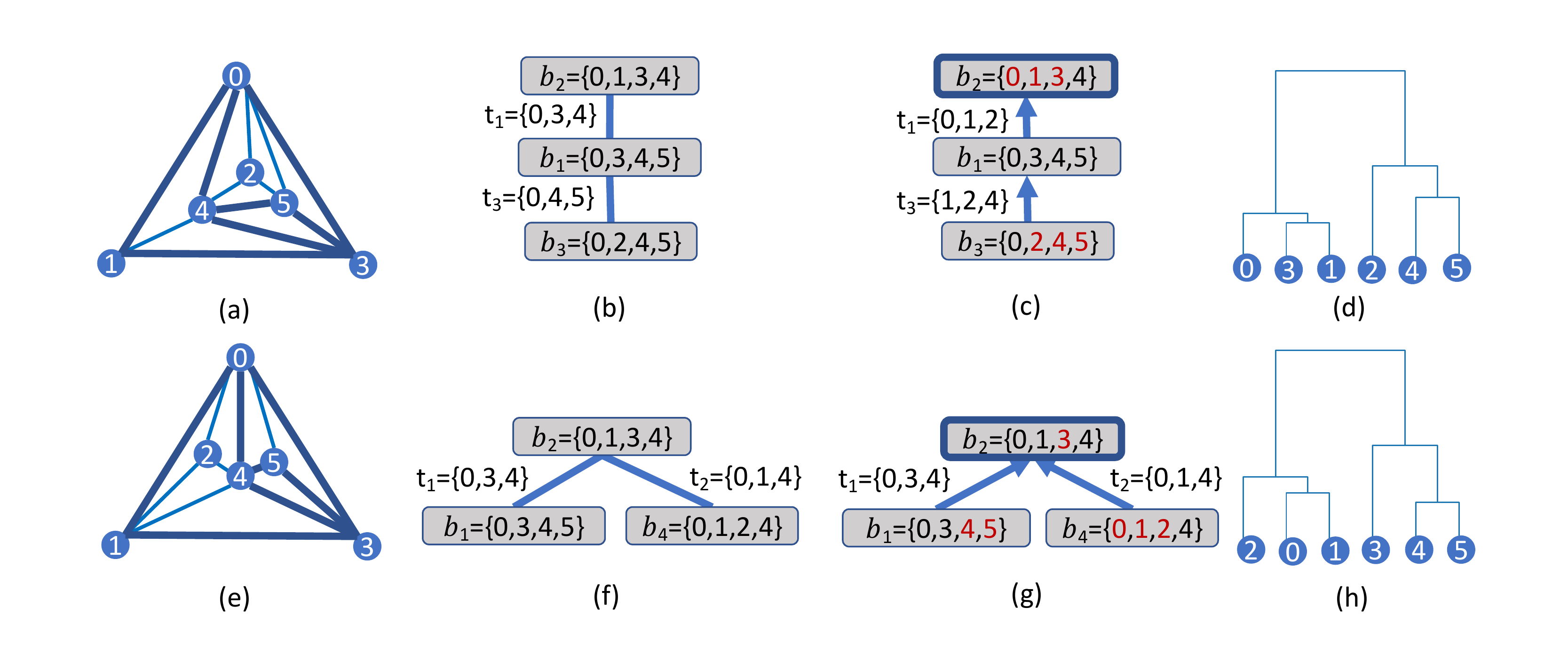}
    \caption{(a)--(d): Example of using \textsc{prefix}=1. (e)--(h): Example of using \textsc{prefix}=3.}
    \label{fig:bubble}
\end{figure}

\showfull{Now we give a more detailed example showing the case when using a larger prefix gives a better clustering result. The correlation between data points in this example is shown in Figure~\ref{fig:heatmap} and the ground truth clustering is the two clusters $\{0,1,2\}$ and $\{3,4,5\}$. We will see that the dendrogram produced using \textsc{prefix}=3 can recover this ground truth clustering, but the dendrogram produced using \textsc{prefix}=1 cannot. Consider the example in Figure~\ref{fig:bubble}. Figure~\ref{fig:bubble}(a) is the TMFG with \textsc{prefix}=1 and Figure~\ref{fig:bubble}(b) is the TMFG with \textsc{prefix}=3. For both prefix sizes, the TMFG algorithm starts with the 4-clique $\{0,1,3,4\}$. When \textsc{prefix}=1, we first add vertex $5$ to $t_1 = \{0,3,4\}$ and then add vertex $2$ to $t_3 = \{0,4,5\}$. On the other hand, when \textsc{prefix}=3, we add vertex $2$ and $5$ in one round. Since $t_3$ does not exist when we add vertex $2$, we add vertex $2$ to $t_2 = \{0,1,4\}$. In Figures~\ref{fig:bubble}(c) and (g), the vertices are assigned to the bubble where they are marked red. 

When \textsc{prefix}=1 we get the dendrogram shown in Figure~\ref{fig:bubble}(d). We see that the ground truth clustering cannot be recovered because vertex $2$ is placed in $t_3$ instead of $t_2$, since $corr(2,5) = 0.42$ is slightly larger than $corr(2,1) = 0.41$, but this difference could be due to noise in the data set. 
When \textsc{prefix}=3 we get the dendrogram shown in Figure~\ref{fig:bubble}(h), and we can obtain the ground truth clustering result by filtering out the noise. }

\fi

\end{document}

%% file: macro.tex
\SetKwInput{KwRequire}{Prereq}
\SetKwInput{KwData}{Input}
\SetKwInput{KwResult}{Output}
\SetKwFor{PFor}{parallel\_for }{do}{end}

\newcommand{\prefix}{\textsc{prefix}}
\newcommand{\clique}{\mathcal{C}}
\newcommand{\bubble}{b}
\newcommand{\faces}{\mathcal{F}}
\newcommand{\outerf}{\text{outer face}\xspace}

\newcommand{\myparagraph}[1]{ \noindent {\bf #1.}}
\newcommand{\defn}[1]{\emph{\textbf{#1}}}

\definecolor{myblue}{RGB}{0,128,255}
\newcommand{\custo}[1]{{\color{myblue} #1}}

\newcommand{\showfull}[1]{{#1}}

\newcommand{\revised}[1]{{#1}}

%% file: abstract.tex
\begin{abstract}
Given all pairwise weights (distances) among a set of objects, filtered graphs provide a sparse representation by only keeping an important subset of weights. Such graphs can be passed to graph clustering algorithms to generate hierarchical clusters. In particular, the directed bubble hierarchical tree (DBHT) algorithm on filtered graphs has been shown to produce good hierarchical clusters for time series data.

We propose a new parallel algorithm for constructing triangulated maximally filtered graphs (TMFG), which produces valid inputs for DBHT, and a scalable parallel algorithm for generating DBHTs that is optimized for TMFG inputs. In addition to parallelizing the original TMFG construction, which has limited parallelism, we also design a new algorithm that inserts multiple vertices on each round to enable more parallelism. We show that the graphs generated by our new algorithm have similar quality compared to the original TMFGs, while being much faster to generate. Our new parallel algorithms for TMFGs and DBHTs are 136--2483x faster than state-of-the-art implementations, while achieving up to 41.56x self-relative speedup on 48 cores with hyper-threading, and achieve better clustering results compared to the standard average-linkage and complete-linkage hierarchical clustering algorithms. We show that on a data set, our algorithms produce clusters that align well with human experts' classification.
\end{abstract}

%% file: intro.tex
\section{Introduction}\label{sec:intro}
Clustering is an unsupervised machine learning method that has been
widely used in many fields including finance,  biology, and computer
vision, to discover structures in a data
set. %
Often times, one is interested in exploring clusters at varying resolutions.
A \emph{hierarchical
  clustering} algorithm can be used to produce a tree, also known as a \emph{dendrogram}, that represents
clusters at different scales.

Running a metric clustering algorithm on a set of $n$ points often involves working with $\Theta(n^2)$ pairwise distances, and is computationally prohibitive on large data sets.
One approach to improving efficiency is to use a \emph{filtered graph} that keeps only a subset of the pairwise distances, and then pass the resulting graph to a graph clustering algorithm.
Filtered graphs
reduce the number of distances considered while retaining the most important
features, both locally and globally.
Simply removing
all edges with weights below a certain threshold may not perform well
in practice, as the threshold may require significant tuning, and the
importance of an edge is not only determined by its weight, but also
its location in the graph.  Several other methods for
constructing filtered graphs have been proposed, including 
$k$-nearest neighbor graphs~\cite{Ruan2010}, minimum
spanning trees~\cite{Mantegna1999,Tumminello2007}, 
and weighted maximal planar graphs~\cite{Tumminello2005, Massara2017}.

A variant of weighted maximal planar graphs, called
planar maximally
filtered graphs (PMFG)~\cite{Tumminello2005},
has been shown to perform well in
practice
in combination with the directed bubble hierarchy tree (DBHT) technique,
especially on financial~\cite{musmeci2015relation, wang2017multiscale, yen2021using} and %
biological~\cite{song2015multiscale,burton2015pathogenesis} %
data sets.
The PMFG is constructed by repeatedly adding an edge between the pair of points with the highest weight, while preserving planarity. The DBHT technique constructs a dendrogram based on certain triangles in the input graph, along with various shortest path calculations.
It has the benefit that no prior 
information about the data is required, and no parameter tuning is needed.
However, the state-of-the-art PMFG and DBHT
algorithms are sequential and do not scale to large data sets.
Massara et. al~\cite{Massara2017} proposed the triangulated
maximally filtered graphs (TMFG), a variant of maximal weighted planar graph that can be generated
more efficiently. Instead of repeatedly adding a single edge as in PMFG, the TMFG is constructed by 
repeatedly adding a single vertex to a triangle with its three edges to the triangle corners, while respecting planarity. 
In other words, we search for an uninserted vertex $x$ that is "close" to three vertices that already form a triangle in the graph, and insert $x$
by adding three edges from $x$ to the three vertices.
However, the state-of-the-art TMFG implementation is also sequential, and  
to the best of our knowledge, the clustering quality of 
using DBHTs with TMFGs has not been evaluated. 

\begin{figure}[!t]
    \centering
    \vspace{-8pt}
    \includegraphics[width = 0.9\columnwidth]{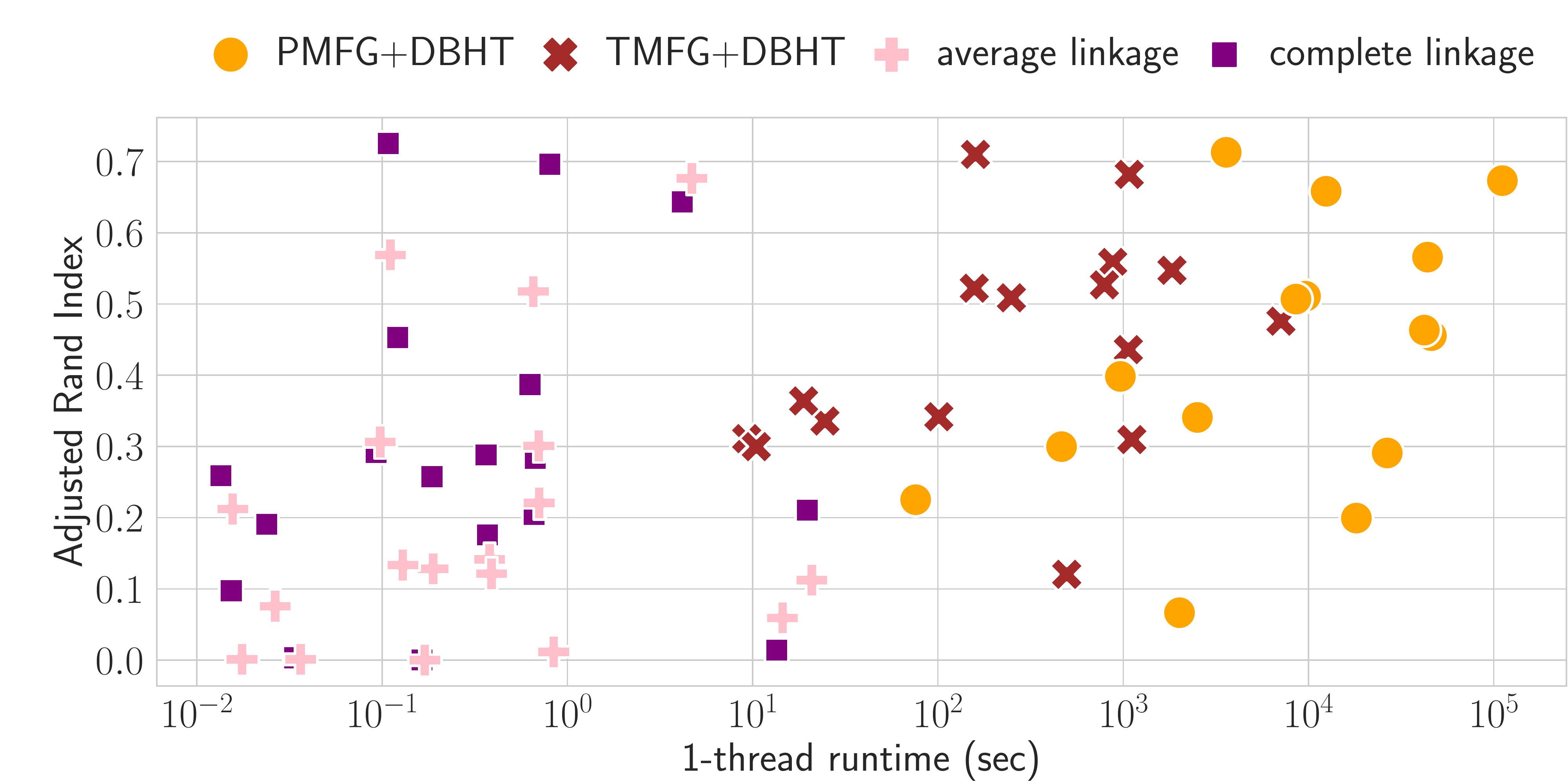}
    \caption{Sequential runtime vs.\ clustering quality for hierarchical clustering methods on different data sets. }
    \label{fig:intro}
\end{figure}

One important type of data that PMFGs, TMFGs, and DBHTs have been shown to perform well on is \emph{time series data}.
Time series data arise in a multitude of applications, including in
finance, signal processing, biology, astronomy, and weather
forecasting. To extract insights from the data, one is often
interested in finding correlations between different time series, and
clustering the data based on these correlations.  The \emph{correlation matrix}
stores the correlation between every pair of time series
It is important to construct a filtered graph on the correlation matrix to enable efficient and scalable clustering.
We show in \cref{fig:intro} the runtime and cluster quality (using the Adjusted Rand Index~\cite{Hubert1985}) for PMFG and TMFG combined with DBHT, compared with the standard average-linkage and complete-linkage methods for hierarchical clustering, for a collection of time series data sets.
For all methods, the dendrogram is cut such that the number of resulting clusters is the same as the number of
ground truth clusters.
We see that, while the runtimes of PMFG and TMFG are higher than those of average-linkage and complete-linkage, they are able to generate higher quality clusters. Thus, PMFG, TMFG, and DBHT are preferable for hierarchical clustering, when one is willing to use a larger time budget to obtain better quality clusters.

To reduce the runtime of TMFG and DBHT, we propose new parallel algorithms and fast
implementations for constructing TMFGs
and a scalable parallel DBHT algorithm
that is optimized for TMFGs. 
Besides parallelizing the original TMFG construction, which has limited parallelism,
we also design a new algorithm that inserts multiple
vertices on each round to enable more parallelism. We show that the filtered graph generated by our new algorithm has similar quality to that of the original TMFG, while being much faster.
The key challenge in our algorithm is in resolving the conflicts when inserting multiple
vertices into the TMFG.
In TMFG construction, we resolve the conflict of vertices by 
designating a single triangle for each vertex based on edge weights.
Our DBHT algorithm leverages the 
special topological structure of TMFGs.
We are able to construct a rooted undirected bubble tree on the fly with a special invariant  while constructing
the TMFG, and we use this invariant to efficiently direct the bubble tree with a recursive process. 
This reduces the complexity of these two steps from quadratic (which the original DBHT algorithm requires) to linear. 

We compare the
speed and quality of our parallel algorithms with parallelized versions of the average-linkage and complete-linkage hierarchical
clustering algorithms. As a baseline, we also compare with $k$-means, which is a non-hierarchical clustering algorithm and only produces clusters at a single resolution. On a collection of 16 data sets generated from time series and image data, we find that 
the DBHT using TMFG produces similar and sometimes better clusters than the DBHT using PMFG.
It also gives better clusters than average-linkage and complete-linkage clustering, and is competitive with $k$-means on most data sets.

We show that our
new algorithm is up to 15589 times faster than the
sequential DBHT on PMFG and
has Adjusted Rand Index scores up to 0.65 higher
than agglomerative clustering algorithms.
We show that
on time series data sets of stock prices from 2013--2019 from the US stock market, DBHT on TMFG is able to produce clusters
that align well with human experts' classification, and in our experiment it  produced better clusters than the original TMFG algorithm. 
On 48 cores with hyper-threading, our new algorithm for constructing a TMFG variant and DBHT are 136--2483x faster times faster than the state-of-the-art TMFG implementation, and achieves up to 41.56x self-relative speedup on 48 cores.

We summarize our contributions below: 
\begin{itemize}[topsep=1pt,itemsep=0pt,parsep=0pt,leftmargin=10pt]
    \item We design a new parallel algorithm for constructing a TMFG variant that produces high-quality graphs.
    \item We design a new parallel algorithm for constructing the DBHT, which is optimized for TMFG-like inputs.
    \item We perform experiments showing that our parallel algorithms achieve significant speedups over state-of-the-art algorithms, while producing clusters of similar or better quality. 
\end{itemize} 
 
Our source code and data is available at \url{https://github.com/yushangdi/par-filtered-graph-clustering}.

%% file: fig_example.tex
\begin{figure*}
\begin{minipage}{\textwidth}
     \centering
    \includegraphics[width = 0.9\textwidth,trim={0 6cm 0 6cm},clip]{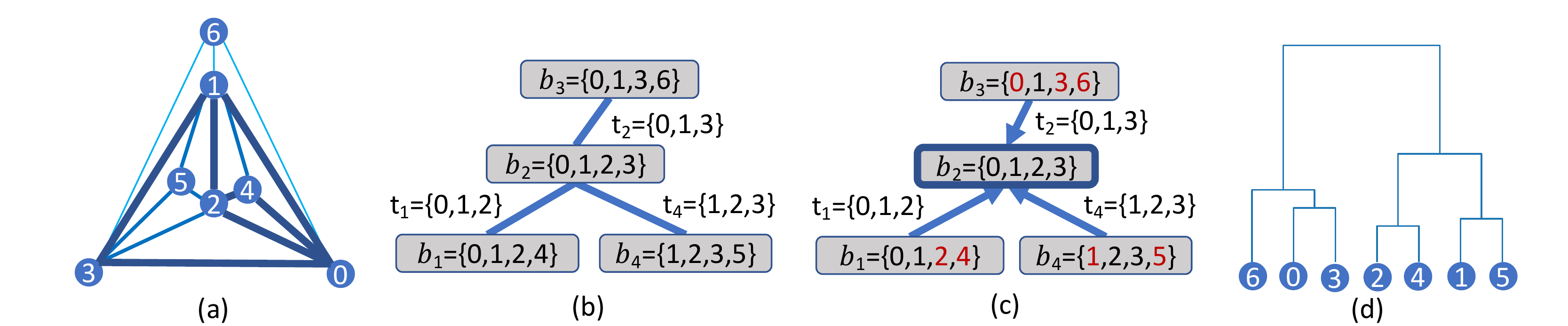}
    \caption[Caption for example]{\textbf{(a)} An example TMFG. The edges have weights $0.8$, $0.4$, or $0.2$. The darker and thicker edges have larger weights. For example, $w(0,1)=0.8$, $w(2,3)=0.4$, and $w(0,6)=0.2$. \textbf{(b)} The undirected bubble tree. Each bubble node is a gray box and nodes are
    connected by blue edges. Each bubble is marked with vertices in the bubble and a name $\bubble_i$. Each edge is marked with the triangle it corresponds to and a name $t_i$. \textbf{(c)} The directed bubble tree. The bolded node is the only converging bubble in this example. The red vertices are assigned to the bubble in which they are marked. 
\textbf{(d)} The dendrogram for DBHT. }%
    \label{fig:example}
\end{minipage}
\end{figure*}

%% file: background.tex
\section{Background}\label{sec:background}

We briefly describe the sequential
construction of PMFGs and TMFGs, and introduce the notation and primitives that we use.

A \defn{planar} graph can be drawn 
on the plane such that no edges cross each other. 
Both PMFGs and TMFGs are maximal planar subgraphs of a
complete undirected, edge-weighted graph ("maximal" means that no additional edge can be added to the vertex set without violating planarity). 
They give an approximation 
to the NP-hard Weighted Maximum Planar Graph problem, which requires the sum of 
the edge weights kept to be maximized~\cite{giffin1984graph}. 
The complete edge-weighted graph input can also be viewed as
a similarity matrix $S$, where $S[i,j]$ is the weight of edge $(i,j)$ in the graph.

\myparagraph{Planar Maximally Filtered Graph (PMFG)}
The sequential algorithm for constructing PMFGs~\cite{Tumminello2005} first
sorts all of the edges by their edge weights.
It then starts with an empty graph $G$ and considers each
edge in the sorted order, from highest to lowest weight. An edge is added
to the graph if and only if it does not violate the planarity of $G$,
which is checked by running a planarity testing algorithm. 
This process is computationally expensive
due to the need to run a planarity testing algorithm $\Theta(n^2)$ times.

\myparagraph{Triangulated Maximally Filtered Graph (TMFG)}
TMFGs~\cite{Massara2017} have
been proposed to improve the efficiency of PMFG construction.  Instead
of considering edges one by one, the original sequential TMFG algorithm adds one
vertex $x$ to the vertex set of the subgraph under construction, as well as three edges (each from $x$ to one of the three vertices of a triangular face in a planar embedding of the subgraph)
on every iteration, and eliminates the vertex from
consideration in future iterations. The vertex-triangle pair that gives
 the highest increase in total edge weight is chosen.
This leads to only $\Theta(n)$
iterations for TMFG, compared to $\Theta(n^2)$ for PMFG. 
The algorithm starts with four vertices with maximum weighted degree and all six edges connecting them.
By definition, it ensures that edges added do not violate 
planarity, and so planarity testing is not needed. 
\cref{fig:example}(a) shows an example of a TMFG. If there were additional vertices to insert, they could go into any of the faces, which are all triangles.

\myparagraph{Directed Bubble Hierarchy Tree (DBHT)} 
After obtaining the TMFG
or PMFG, we can then generate a dendrogram from it for 
hierarchical clustering
using the DBHT~\cite{song2012} algorithm, which
has been shown to produce high-quality dendrograms for financial time
series~\cite{musmeci2015relation,wang2017multiscale, yen2021using} and biological~\cite{song2015multiscale,burton2015pathogenesis} %
data. 
The original, sequential DBHT algorithm first constructs a
\emph{bubble tree}~\cite{song2011nested}, which is a tree where nodes correspond to planar subgraphs, 
and edges between two nodes correspond to triangles in the original graph that separate the two corresponding planar subgraphs.
The DBHT algorithm
then directs the tree edges by computing the strength of connection
from the separating face to its interior and exterior.
Next, the algorithm
runs a number of shortest distance computations to assign vertices to
bubbles, and finally uses complete-linkage clustering to generate a
hierarchy.
We will describe more details of the DBHT algorithm along with 
how we parallelize and optimize it in~\cref{sec:pardbht}.

\myparagraph{Notation and Terminology}
Given a planar graph $G$ and its planar embedding, a \defn{face} of $G$ is a maximal region bounded by edges, except for the \defn{\outerf}, which is unbounded.
For example, 
in \cref{fig:example}(a), $\{0,3,6\}$ is the \outerf in this embedding. 
A bounded face is an \defn{inner face}.
We \defn{insert} vertex $v$ into a triangular face $t$ with three corners  $v_x$, $v_y$, and $v_z$
by adding three weighted edges from $v$ to each of $v_x$, $v_y$, and $v_z$. %
The \defn{gain} of inserting $v$ to $t$ is the sum of these three edge weights.
Let $V$ be a set of vertices. We say that a vertex $v \in V$ is the \defn{best vertex in $V$} for face $t$ if 
inserting $v$ to $t$ gives the maximum gain among all vertices in $V$. 
The weight of an edge $(i,j)$ in graph $G$ is denoted as $w(i,j)$.

A face is \defn{separating} if removing it disconnects the graph.
A \defn{bubble} is a maximal planar graph whose 3-cliques are non-separating cycles.
If we create a \defn{bubble node} for each bubble and connect bubble nodes that 
share separating triangles, then the bubble nodes and connection edges form a tree~\cite{song2011nested, song2012}, which we call a \defn{bubble tree}.
We will denote a clique with $\clique$ 
and a bubble node in the bubble tree as $b$.

The output of our algorithm is a tree called a
\defn{dendrogram}, where the height of each node corresponds to the
dissimilarity between the two clusters represented by its two children. 
The \defn{height} of a dendrogram node
is 
a value computed by our algorithm, and
a node's height is at most the height of its parent.
Cutting the dendrogram at different heights gives subtrees that correspond to
clusters at different scales.

\myparagraph{Model of Computation} To analyze our parallel algorithms, we
  use the \defn{work-span model}~\cite{JaJa92,CLRS}, a standard model for analyzing shared-memory algorithms.  
The work $W$ of an algorithm is the total number of operations, and the span $S$ is the length of the longest dependency path of the algorithm. The work of a sequential algorithm is the same as its time. 
We can bound the running time of the algorithm on $P$ processors by $W/P + O(S)$ using a randomized work-stealing scheduler~\cite{Blumofe1999}.

\myparagraph{Parallel Primitives~\cite{JaJa92,Vishkin10}}
\revised{ \Cref{tab:prelims} gives a summary of the parallel primitives that we use.}
A parallel \defn{filter} takes an array $A$ and a
predicate function $f$, and returns a new array containing $a \in A$
for which $f(a)$ is true, in the same order that they appear in $A$.
Filter takes $O(n)$ work and $O(\log n)$ span.
A parallel \defn{sort} takes an array $A$ and a
binary function $f(a,b)$ that returns true if $a<_kb$ where $<_k$ is a total ordering, and returns a new array containing $a \in A$
in non-decreasing order. Sorting takes $O(n\log n)$ work and $O(\log n)$ span. A parallel \defn{integer sort} can be done in
$O(n)$ work and $O(\log n)$ span with high probability (w.h.p.)\footnote{We
  say that a bound holds \defn{with high probability (w.h.p.)} if it holds with probability at least $1-n^{-q}$ for $q \geq 1$,
  where $n$ is the input size.} for integers in the range $[1,\ldots,O(n\log^c n)]$~\cite{RR89}.
A parallel \defn{maximum} takes an array $A$ and returns the largest element in $A$.
A maximum can be computed in $O(n)$ work and $O(1)$ span with high probability.
\textsc{WriteMin} is a priority concurrent write that takes as input two
arguments, where the first argument is the location to write to and the
second argument is the value to write; on concurrent writes,
the smallest value is written~\cite{ShunBFG2013}. \textsc{WriteMax} is similar but writes the largest value. 
\textsc{WriteAdd} is a priority concurrent write that takes as input two
arguments, where the first argument is the location to write to and the
second argument is the value to atomically add to the value at the first location. 
We assume that these priority concurrent writes take constant work and span.

\begin{table}
\footnotesize
\centering
  \caption{\revised{Parallel Primitives.}}
  \label{tab:prelims}
\begin{tabular}{cccc}
\toprule
Operator & Description & Work & Span  \\ \hline
Filter & Filter out desired elements & $O(n)$ & $O(\log n)$ \\ 
Sort & Sort elements & $O(n)$ & $O(\log n)$ w.h.p. \\ 
Maximum & Find the maximum & $O(n)$ & $O(1)$ w.h.p. \\ 
\textsc{WriteMin} & concurrently write  & $O(1)$ & $O(1)$ \\ 
                   & the smallest value & & \\
\textsc{WriteMax} & concurrently write  & $O(1)$ & $O(1)$ \\ 
                   & the largest value & & \\
\textsc{WriteAdd} & atomically add & $O(1)$ & $O(1)$ \\ 
\bottomrule
\end{tabular}
\end{table}

%% file: related.tex
\section{Related Work}

\revised{
There is a rich literature in designing both sequential and parallel hierarchical
clustering algorithms. %
The most popular hierarchical clustering algorithm variants are 
hierarchical agglomerative clustering algorithms (HAC)~\cite{mullner2013fastcluster, Yu2021, wang2021fast, dhulipala2021hierarchical, rajasekaran2005efficient}. Different variants of HAC use different linkage functions to
compute the distance between clusters. Popular linkage functions
include complete, average, single, centroid, and median linkage.
Several papers have been written on the topic of parallel HAC, such as parallel
HAC under the single-linkage metric~\cite{olson1995parallel, hendrix2012parallel, wang2021fast, Yu2021, dahlhaus2000parallel} as well as other metrics~\cite{Yu2021}.
Song et al.~\cite{song2012} showed that DBHT with PMFG is superior to complete-linkage and average-linkage HAC.
We also show in \Cref{sec:exp} that our method has higher quality than complete-linkage and average-linkage HAC. 
Musmeci et al.~\cite{musmeci2015relation} showed that DBHT with PMFG produces better clusters on stock data sets than single linkage, average linkage, complete linkage, and $k$-medoids. 

There has also been work on
other hierarchical clustering methods, such as partitioning
hierarchical clustering algorithms, algorithms that combine
agglomerative and partitioning
methods~\cite{dash2004efficient,  mao2015parallel}, %
and density-based methods~\cite{Campello2015, wang2021fast}.
Compared to the density-based methods~\cite{Campello2015, wang2021fast}, our algorithm has two advantages, the first is that many density-based methods are most suitable for low-dimensional data, while our algorithm does not have this constraint. Moreover, these clustering methods require some apriori information to set 
the parameters to obtain high-quality clusters. In comparison, DBHT does not 
require any additional parameters.
Dash et al.~\cite{dash2004efficient} proposed a parallel 
hierarchical clustering algorithm that divides data into overlapping cells and clusters the cells. However, the parallel version is only described for low-dimensional data sets. Cohen-Addad et al.~\cite{cohen2019hierarchical} designed a parallel algorithm for the hierarchical $k$-median problem, but it is limited to the Euclidean metric. 
There are also algorithms designed for specific types of data, such as categorical data~\cite{wang2011parallel} and genomics data~\cite{mao2015parallel}. While most parallel algorithms have been designed for shared memory, there have also been algorithms designed for distributed memory~\cite{lattanzi2020framework}.

}

The PMFG is an approximate solution to the weighted maximal planar graph (WMPG) problem, which is NP-hard~\cite{giffin1984graph}. 
There have also been other approximate solutions designed for the WMPG.
Massara et al.~\cite{Massara2017} propose the TMFG, which is based on local topological moves. 
Kataki et al.~\cite{kataki2020new} design a heuristic 
that is based on a transformation to the connected spanning subgraph problem and the
dual graph. Osman et al.~\cite{osman2003greedy} use a greedy random adaptive search procedure 
to solve the WMPG problem. 
Other heuristics have been developed for the WMPG problem~\cite{dyer1985analysis,Cimikowski1995,eades1982efficient}.

Besides filtering graphs based on topological constraints,
there are other
graph filtering methods that are used for clustering. 
For example, Ma et al.~\cite{ma2020towards} use a low-pass filter to extract useful data representations for 
subspace clustering, and
Tremblay et al.~\cite{tremblay2016accelerated}
use graph filtering for faster spectral clustering.
Another graph construction algorithm used for clustering 
is $b$-matching~\cite{jebara2006b, jebara2009graph, emek2021hierarchical}.
The $b$-matching algorithm filters graph edges such
that the in-degree and out-degree of each node is at most $b$, and can be solved in polynomial time.

%% file: tmfg.tex
\section{Parallel TMFG}\label{sec:partmfg}

This section introduces our algorithm for parallelizing TMFG construction, whose pseudocode is shown in \cref{alg:partmfg}.
We first give a high-level overview. Our parallel algorithm simultaneously adds
multiple vertices to the graph.
Adding multiple vertices may give results that
differ from the sequential TMFG algorithm because the sequential
algorithm updates the graph after every insertion, giving future
vertices more faces to choose from, whereas our
parallel algorithm will only update the graph after all of the vertices in
a round have been added. Therefore, the more vertices we add in
parallel, the more we deviate from the sequential algorithm.
This gives a tradeoff in the quality of the graph and the
amount of available parallelism.  We therefore consider adding only a prefix of
the vertices that give the highest gain in total edge weight.  The
prefix size can be tuned for the application and available parallelism.

\input{algo_tmfg}

In the rest of the section, we describe the details of our parallel TMFG algorithm 
(\cref{alg:partmfg}). The input is an $n \times n$ symmetric matrix $S$ that represents the complete 
undirected graph, and a parameter $\prefix$.
The highlighted lines are used for building DBHT, and will be explained in \cref{sec:pardbht}. 
On Lines~\ref{alg:partmfg:first4}--\ref{alg:partmfg:restv}, we first find the four vertices $\{v_1, v_2, v_3, v_4\}$ that have the highest total sum
across their rows in $S$, and add all edges among them to the 
resulting graph's edge list $\mathcal{E}$. The four faces formed by the four vertices are added to the set of faces $\faces$.
The rest of the vertices are in set $V$, and will be added to the graph later. 
On Line~\ref{alg:partmfg:initgain}, for each face in $\faces$, we find its best vertex (the vertex that maximizes the gain if inserted into this face).

On Lines~\ref{alg:partmfg:while}--\ref{alg:partmfg:return}, we insert the remaining vertices into the graph in batches of up to size $\prefix$. 
On Lines~\ref{alg:partmfg:getprefix}--\ref{alg:partmfg:restv2}, we obtain the batch of vertices to insert and their corresponding faces to insert into. 
We first obtain the $\prefix$ vertex-face pairs with the largest gains in the $\textsc{Gains}$ array, and store the pairs in $L$. 
This can be done using a parallel sort on the $\textsc{Gains}$ array. 
On Line~\ref{alg:partmfg:conflict}, we ensure that a given vertex is only added to a single face to avoid race conditions. Conflicts for a given
vertex are resolved by only allowing the vertex to add itself to the face that gives the highest gain.
We can use
parallel sorting to group vertices based on which face (among all
faces) gives the highest gain, and have each face pick the vertex
(among all vertices that choose this face) that gives the highest gain.
The available parallelism increases as the number of faces in the
graph grows. 
If $\prefix=1$,  Lines~\ref{alg:partmfg:getprefix}--\ref{alg:partmfg:conflict} can be simplified to a single parallel maximum computation. 
On Line~\ref{alg:partmfg:restv2}, we remove the vertices in $L$ from $V$ using a parallel filter.

On Lines~\ref{alg:partmfg:for}--\ref{alg:partmfg:updategainloop}, we loop over the vertex-face pairs $(v,t)$ in $L$, add $v$ to $t$, and 
update $\mathcal{E}$ and $\faces$. To update $\faces$, we add the three new faces created and remove $t$. 
We also update $\textsc{Gains}$ by computing the new best vertex among $V$ for the three new faces and faces that previously had $v$ as their best vertex. 
Unlike the original algorithm~\cite{Massara2017}, which
loops over all of the faces to find the faces that previously had $v$ as their best vertex, we keep an array for each vertex that stores such faces to improve the efficiency of this step.

If we choose $\prefix=1$ in \cref{alg:partmfg}, we obtain the 
same result as the sequential TMFG algorithm, but we still have some parallelism 
from the parallel maximum call on Line~\ref{alg:partmfg:getprefix}
on Lines~\ref{alg:partmfg:updategain}--\ref{alg:partmfg:updategainloop}. 
This case gives the same parallel algorithm that is described (but not implemented) by Massara et al.~\cite{Massara2017}. However,
we show in~\cref{sec:exp} that using $\prefix=1$ has limited parallelism 
because on each round only a single vertex can be inserted.

\input{algo_bubble}

%% file: algo_tmfg.tex
\begin{algorithm}[!t]
\DontPrintSemicolon
\fontsize{7pt}{7pt}\selectfont
\KwData{$n \times n$ similarity matrix $S$, prefix size \prefix \ $\geq 1$}
\tcp*[h]{The four vertices that have highest total sum across its row in $S$}\;
$\clique = \{v_1, v_2, v_3,  v_4\}$\;  \label{alg:partmfg:first4}
$\mathcal{E} = \{(v_1, v_2)
(v_1, v_3),
(v_1, v_4),
(v_2, v_3),
(v_2, v_4),
(v_3, v_4)\}$\;
\tcp*[h]{The four triangular faces in the initial graph}\;
$\faces  =\{\{v_1,v_2,v_3\},\{v_1,v_2,v_4\},\{v_1,v_3,v_4\},\{v_2,v_3,v_4\}\}$\; %
$V = \{v_5, \dots, v_n\}$\; \label{alg:partmfg:restv}
\tcp*[h]{The best vertex in $V$ for each triangle}\;
$\textsc{Gains} = [\operatorname*{argmax}_{u \in V}(\sum_{v \in t} S[v,u])$ for $t\in \faces ]$ \; \label{alg:partmfg:initgain}
\custo{Initialize bubble tree $T$ with $\clique$}\; \label{alg:partmfg:initbbtree}%
\custo{$\textsc{OuterFace}=\{v_1, v_2, v_3\}$}\;  \label{alg:partmfg:initouterface}
\While{$|V|>0$}{\label{alg:partmfg:while}
    $L$ = the $\prefix$ vertex-face pairs with the largest gains in $\textsc{Gains}$ using parallel sorting.\; \label{alg:partmfg:getprefix}
    For vertices paired with multiple faces in $L$, only keep the pair with maximum gain, using a parallel filter and parallel sorting.\; \label{alg:partmfg:conflict}
    $V = V \setminus \{\text{vertices} \in L\}$\; \label{alg:partmfg:restv2}
    \PFor{$(v,t= \{v_x, v_y, v_z\}) \in L$ }{\label{alg:partmfg:for}
        $\mathcal{E} =\mathcal{E} \cup \{(v, v_x), (v, v_y), (v, v_z)\}$\;\label{alg:partmfg:13}
        $\faces = \faces \cup \{\{v, v_x, v_y\}, \{v, v_y, v_z\}, \{v, v_x, v_z\}\} \setminus t $\;\label{alg:partmfg:14}
        \PFor{$t \in$ $\{t': \textsc{Gains} [t']=v\} \cup  \{\{v, v_x, v_y\}, \{v, v_y, v_z\}, \{v, v_x, v_z\}\}$ }{ \label{alg:partmfg:updategain}
            $\textsc{Gains}[t]$ = $ \operatorname*{argmax}_{u \in V}(\sum_{v \in t} S[v,u])$ \; \label{alg:partmfg:updategainloop}
        }
        \custo{UpdateBubbleTree($v,t, T$)}\; \label{alg:partmfg:updatebubble}
    }
}
\Return $\mathcal{E}$\; \label{alg:partmfg:return}
\caption{Parallel TMFG}\label{alg:partmfg}
\end{algorithm}

%% file: algo_bubble.tex
\begin{algorithm}[!t]
\DontPrintSemicolon
\fontsize{7pt}{7pt}\selectfont
  \SetKwFunction{FMain}{UpdateBubbleTree}
  \SetKwProg{Pn}{Function}{:}{}
  \Pn{\FMain{$v$, $t= \{v_x, v_y, v_z\}$, $T$}}{
    $\bubble^*$ = new bubble $\{v, v_x, v_y, v_z\}$\;\label{alg:partmfg:bbnew}
    $\bubble$ = the bubble that $t$ is in\; \label{alg:partmfg:bbt}
    \eIf{$t =\textsc{OuterFace}$}{ \label{alg:partmfg:outerif}
        $\bubble$.\textsc{parent} = $\bubble^*$\; 
        Add $\bubble$ to $\bubble^*$.\textsc{children}\;
        $\textsc{OuterFace} = \{v, v_x, v_y\}$\; \label{alg:partmfg:setouter}
    }{ \label{alg:partmfg:innerif}
        $\bubble^*$.\textsc{parent} = $\bubble$\;
        Add $\bubble^*$ to $\bubble$.\textsc{children}\; \label{alg:partmfg:innerend}
    }
  }
\caption{UpdateBubbleTree}\label{alg:updatebubbletree}
\end{algorithm}

%% file: dbht.tex
\section{Parallel DBHT for TMFG}\label{sec:pardbht}
We describe our parallel algorithm for building the DBHT, which leverages the special structure of the bubble tree
for TMFGs. 
The original DBHT construction of Song et al.~\cite{song2012} takes a planar graph,
and two weights on each edge---a similarity 
measure and a dissimilarity measure. The similarity measure is 
from the similarity matrix $S$, and the dissimilarity measure needs to be additionally supplied. 
The construction has the following steps:
building an undirected 
bubble tree~\cite{song2011nested} from a planar graph;
assigning directions to the bubble tree edges;
assigning graph vertices to bubbles;
and using complete-linkage clustering to generate a
hierarchy. 
Below, we describe how we accelerate and parallelize each step.

\subsection{Bubble Tree for TMFG}
As described in ~\cref{sec:background}, a 
\emph{bubble tree}~\cite{song2011nested} is a tree where nodes correspond to planar subgraphs, 
and edges between nodes correspond to triangles in the original graph that separate the two corresponding planar subgraphs.
Specifically, a \emph{bubble} is a maximal planar graph whose triangles are non-separating~\cite{song2011nested}.
The original DBHT construction is inefficient,
as it involves first finding all of the triangles in the graph,
and then testing for every triangle whether removing it would
disconnect the graph. 

The key observation is that the bubble tree can be more
efficiently constructed during TMFG construction, as a vertex that is
added in the TMFG algorithm will correspond to exactly one bubble node and one edge in the bubble tree.
In TMFG, every time we insert a new vertex, we create a 4-clique,
which is a bubble because all faces of a 4-clique are non-separating triangles. 
This will also produce a separating triangle, which corresponds to a new edge in the bubble tree. 
This separating triangle is shared by the 4-cliques corresponding to the two bubble nodes incident to it.
For example, consider the graph without vertex 5 in \cref{fig:example}(a). If we now insert vertex 5 into 
triangle $t_4 = \{1,2,3\}$, we created a new bubble $b_4 = \{1,2,3,5\}$ and $t_4$ now becomes a separating triangle, which
is shared by $b_2=\{0,1,2,3\}$ and $b_4$.
We
incorporate the bubble tree construction into our parallel TMFG
algorithm to save a significant amount of work. 

Our undirected bubble tree for TMFG has the invariant that each bubble node has a parent and at most
three children (because for each 4-clique, there is one outer face and three inner faces), except for the root, which does not have a parent. 
Moreover, all descendants of an edge are on the interior side of the separating triangle corresponding to the edge.
This invariant is important for us to accelerate the direction computation,
which we describe later.

We now describe the details of bubble tree construction, which are
the highlighted lines in \cref{alg:partmfg}.
On Line~\ref{alg:partmfg:initbbtree}, we initialize our bubble tree with 
a single bubble node, which contains the 4-clique that we start our TMFG with.
On Line~\ref{alg:partmfg:initouterface}, we initialize our outer face to be
$\{v_1,v_2,v_3\}$. This outer face can be chosen arbitrarily 
among the four faces of $\clique$
because the bubble tree's topological
structure is independent of this choice~\cite{song2011nested}.
On Line~\ref{alg:partmfg:updatebubble}, we add a new node 
and a new edge to the bubble tree $T$ for each vertex $v$
inserted into face $t$. This can be done in parallel because
each vertex $v$ is inserted into a single, unique face $t$, 
so there will not be any conflicts.

\cref{alg:updatebubbletree} shows the subroutine for building the bubble tree.
On Line~\ref{alg:partmfg:bbnew}, we create a new bubble node $\bubble^*$.
On Line~\ref{alg:partmfg:bbt}, we find the bubble $\bubble$ in $T$ 
that $t$ is in. This is the bubble that is created when $t$ is created (in some previous call to UpdateBubbleTree). %
$\bubble^*$ is the bubble that the faces $\{v, v_x, v_y\}, \{v, v_y, v_z\}$, and $\{v, v_x, v_z\}$
are in. 
If $t$ is the current outer face, this means we are inserting $v$ into the 
outer face, and $\bubble$ is the current root of $T$. So on Lines~\ref{alg:partmfg:outerif}--\ref{alg:partmfg:setouter}, we let $\bubble$'s parent
be $\bubble^*$, and add $\bubble$ to $\bubble^*$'s children. This maintains the invariant above because the vertices in the bubbles before inserting $v$ are in the interior of the outer face, and after the procedure these bubbles become descendants of the edge corresponding to the outer face $t$.
The $\textsc{OuterFace}$ 
is then updated to be a face $\{v,v_x,v_y\}$ in the 4-clique corresponding to $\bubble^*$.
If $t$ is not the current outer face, then we do not need to change the
outer face, because the vertex is inserted into some inner face of $\bubble$. In this case, on Lines~\ref{alg:partmfg:innerif}--\ref{alg:partmfg:innerend}, we let $\bubble^*$'s parent
be $\bubble$, and add $\bubble^*$ to $\bubble$'s children.
This maintains the invariant above because $v$ must be in the interior of $t$ and its ancestors, and the bubble $\bubble^*$ containing $v$ is a descendant of the edge corresponding to $t$.

\myparagraph{\revised{Example 1}} We now give an example of the bubble tree construction by running our algorithm on the example TMFG in \cref{fig:example}.
We will first describe inserting a single vertex, and then 
describe inserting multiple vertices in parallel.
Suppose we start with the 4-clique  $\clique=\{0,1,2,4\}$.
We have four faces $\{\{0,1,2\}$, $\{0,1,4\}$, $\{0,2,4\}$, $\{1,2,4\}\}$,
where $t_1 = \{0,1,2\}$ is the outer face.
We initialize the bubble tree $T$ with node $b_1=\{0,1,2,4\}$,
which corresponds to $\clique$.
We insert vertices $3$, $5$, and $6$, in order,
into faces $\{0,1,2\}$, $\{1,2,3\}$, and $\{0,1,3\}$, respectively.
We first insert $3$ into $t=t_1=\{0,1,2\}$.
For this insertion, the new bubble
$\bubble^*$ on Line~\ref{alg:partmfg:bbnew}
is $b_2=\{0,1,2,3\}$, and the $b$ on Line~\ref{alg:partmfg:bbt} is $b_1$. 
Since $t$ is the current outer face,
we let $b_2$ be $b_1$'s parent.
The new outer face is $t_2=\{0,1,3\}$.
Now suppose we insert $5$ and $6$
into $\{1,2,3\}$ and $\{0,1,3\}$, respectively, in parallel.
This is similar to inserting $3$, and we
add $b_3=\{0,1,3,6\}$ as $b_2$'s parent and $b_4=\{1,2,3,5\}$ as $b_2$'s child in parallel.

\subsection{Directing Bubble Tree Edges}
We now describe how to direct the bubble tree edges after obtaining the 
undirected bubble tree from Algorithms \ref{alg:partmfg} and \ref{alg:updatebubbletree}, 
and how we significant improve the efficiency of this step over the original sequential DBHT algorithm.
Each edge of the bubble tree corresponds to a separating triangle.
The direction is decided by computing the sum over the weights of the edges in the 
TMFG connecting the triangle with its interior versus its exterior~\cite{song2012}. \cref{fig:example}(c) shows an example of directing the edges.

We denote the sum over the weights of the edges to the interior as \textsc{inVal} and to the exterior 
as \textsc{outVal}.
In the original algorithm, the two values are computed by running a 
breadth-first-search (BFS) on $G \setminus t$ for each separating triangle $t$ to find its exterior and
interior, and then computing the sum of edge weights. This takes $\Theta(n^2)$ work because PMFGs and TMFGs
have $\Theta(n)$ edges~\cite{song2012} and each BFS takes $\Theta(n)$ work. 

In our bubble tree, the interior of a separating triangle contains all of the vertices in the descendants of the edge corresponding to this separating triangle,
and the exterior contains all other vertices. Using this property,
we can compute the direction of all edges in $\Theta(n)$ work using a novel recursive algorithm, starting from the root of bubble tree. At each bubble node $\bubble$, we compute the direction of the edge from itself to its parent, and recursively call the procedure on the bubble's children. 
This gives a significant improvement over the quadratic work of the original algorithm.

\begin{algorithm}[!t]
\DontPrintSemicolon
\fontsize{7pt}{7pt}\selectfont
\SetKwFunction{FMain}{ComputeDirection}
  \SetKwProg{Pn}{Function}{:}{}
  \Pn{\FMain{\text{bubble tree node} $\bubble$}}{
\eIf{$\bubble$ has a parent}{
$\{v_x, v_y, v_z\}$ = vertices shared by $\bubble$ and its parent\; \label{alg:direction:gettri}
$v =  \bubble \setminus \{v_x, v_y, v_z\}$\; \label{alg:direction:getv}
$r = \{\}$\;\label{alg:direction:initr}
$r[v_x]=w(v_x,v), r[v_y]=w(v_y,v), r[v_z]=w(v_z,v)$\;\label{alg:direction:initr2}
\PFor{$\bubble^* \in \bubble$'s children }{  \label{alg:direction:loop}
    $r^*=\{v_x^*:val_x^*, v_y^*:val_y^*, v_z^*:val_z^*\}$ = computeDirection($\bubble^*$)\; \label{alg:direction:loopchild}
    \lIf{$v_x^* \in r$}{\textsc{WriteAdd}($r[v_x^*], val_x^*$)}\label{alg:direction:loop1}
    \lIf{$v_y^* \in r$}{\textsc{WriteAdd}($r[v_y^*], val_y^*$)}
    \lIf{$v_z^* \in r$}{\textsc{WriteAdd}($r[v_z^*], val_z^*$)} \label{alg:direction:loop2}
}
$\textsc{inVal} = r[v_x]+r[v_y]+r[v_z]$ \;\label{alg:direction:inval}
$\textsc{outVal}=deg(v_x)+deg(v_y)+deg(v_z)-\textsc{inVal}-2(w(v_x, v_y)+w(v_x, v_z)+w(v_y, v_z))$\;\label{alg:direction:outval}
\eIf{$\textsc{inVal}>\textsc{outVal}$}{ \label{alg:direction:compare}
    Direct the edge from $\bubble$'s parent to $\bubble$\;
}{
    Direct the edge from $\bubble$ to $\bubble$'s parent\; \label{alg:direction:compare2}
}
\Return $r$\;  \label{alg:direction:return}
}(\tcc*[h]{$\bubble$ is the root}){
\PFor{$\bubble^* \in \bubble$'s children }{ \label{alg:direction:root1}
    computeDirection($\bubble^*$)\;
}
\Return $\{\}$\;\label{alg:direction:root2}
}
}
\caption{ComputeDirection}\label{alg:direction}
\end{algorithm}

\input{algo_dbht}

Specifically, we compute the direction of all edges by recursively calling the \textsc{ComputeDirection} function, shown in \cref{alg:direction}, with the root of bubble tree as the argument to the initial call. 
At each bubble node $\bubble$, we compute the direction of the edge from itself to its parent.
If $\bubble$ is the root, then it has no edge to its parent, and so we do not need to compute anything besides initializing the computation on its children (Lines~\ref{alg:direction:root1}--\ref{alg:direction:root2}). 
Otherwise, we compute the \textsc{inVal} and \textsc{outVal} for the separating triangle corresponding to the edge from the bubble to its parent. If $\textsc{inVal}>\textsc{outVal}$, then the
edge goes from $\bubble$'s parent to $\bubble$, and vice versa (Lines~\ref{alg:direction:compare}--\ref{alg:direction:compare2}). 

On Lines~\ref{alg:direction:gettri}--\ref{alg:direction:getv}, we obtain the vertices $\{v_x, v_y, v_z\}$ in the separating triangle represented by the edge from $\bubble$ to its parent, 
and let $v$ be the remaining vertex in the bubble.
On Lines~\ref{alg:direction:initr}--\ref{alg:direction:initr2}, we initialize the sum of edges to the interior from each corner of the triangle with
the edge weights from each corner to $v$ since $v$ is in the interior.
Then, on Lines~\ref{alg:direction:loop}--\ref{alg:direction:loop2} we recursively, and in parallel, compute the sum of edge weights
from the corners of $\bubble$'s children to the children's interior. Note that this computation is nested parallel, meaning that parallel tasks are recursively created.
Since the children's interior is also $\bubble$'s interior, we can sum the weights of $\bubble$'s corners
obtained from its children to compute the \textsc{inVal} of $\bubble$. 
On Line~\ref{alg:direction:outval}, we compute \textsc{outVal} by subtracting the \textsc{inVal}
and the edge weights of the triangle from the weighted degrees of the corners of the triangle. 
On Line~\ref{alg:direction:return}, we return the weights at the corners to $\bubble$'s parent.

Now, we describe the derivation of the formula on Line~\ref{alg:direction:outval} of \cref{alg:direction}.
Let $D = deg(v_x)+ deg(v_y) + deg(v_z) $, $I_k$ be the total weighted degree of $v_k$ to the interior, and $O_k$ be the total weighted degree of $v_k$ to the exterior. We can rewrite $D$ as follows:
\begin{align*}
    D&= I_x +  O_x + w(v_x,v_y)+w(v_x,v_z) \\ 
    &+I_y + O_y + w(v_y,v_x)+w(v_y,v_z)\\
    &+ I_z +  O_z + w(v_z,v_x)+w(v_z,v_y)\\
    &= (I_x+ I_y + I_z) + (O_x  + O_y + O_z)\\
    &+2 (w(v_x,v_y)+w(v_x,v_z)+w(v_y,v_z))\\
    &=\textsc{inVal} + \textsc{outVal} +2 (w(v_x,v_y)+w(v_x,v_z)+w(v_y,v_z))
\end{align*}

Rearranging gives the following formula for computing \textsc{outVal}:
$     \textsc{outVal}  =  D - \textsc{inVal} - 2 (w(v_x,v_y)+w(v_x,v_z)+w(v_y,v_z))$

\myparagraph{\revised{Example 2}} We continue to our example for \cref{alg:direction}. Given the rooted bubble tree in \cref{fig:example}(b),
we start our computation at the root $b_3$.
Since $b_3$ does not have a parent, we recurse down to
its child $b_2$. For $\bubble = b_2$, the vertices shared with its parent are
$\{v_x,v_y,v_z\} = t_2 = \{0,1,3\}$ and the remaining vertex is
$v=2$. Then, we initialize $r[0] = w(0,2)$,
$r[1] = w(1,2)$, and $r[3] = w(3,2)$ on \cref{alg:direction:initr2}.
Next, we recurse down to $b_2$'s children, $b_1$ 
and $b_4$. For $b_1$, the resulting $r^*$ array should contain 
$r^*[0]=w(0,4)$, $r^*[1]=w(1,4)$, and $r^*[2]=w(2,4)$ from the recursion on Line~\ref{alg:direction:loopchild}
because the shared vertices with its parent are $t_1 = \{0,1,2\}$ and the remaining vertex is $4$.
On Lines~\ref{alg:direction:loop1}--\ref{alg:direction:loop2},
since $v_x^*=0\in r$ and $v_y^*=1\in r$,
we increment $r[0]$ by $w(0,4)$
and $r[1]$ by $w(1,4)$.
Since $v_z^* = 2 \notin r$, we do not process it.
Now $r[0] = w(0,4) + w(0,2)$ and $r[1]=w(1,4) + w(1,2)$.
Similarly for $b_4$,  we increment $r[1]$ by $w(1,5)$
and $r[3]$ by $w(3,5)$.
Now $r[0] = w(0,4) + w(0,2)$, $r[1]=w(1,4) + w(1,2)+w(1,5)$ and $r[3]=w(3,2)+w(3,5)$.

When we get to Line~\ref{alg:direction:inval} for $b_2$,
the sum of $r[0]$, $r[1]$, and $r[3]$ is \textsc{inVal} because it contains
the edge weights from the three corners of $t_2$
to its interior.  \textsc{outVal} is computed by 
summing the weights of all edges from $t_2$ and
then subtracting \textsc{inVal} and the weight of $t_2$'s edges (once in each direction).
In our example, we find that $\textsc{inVal}>\textsc{outVal}$, and so 
the edge is directed from $b_3$ to $b_2$ (\cref{fig:example}(c)).
The other edges are directed similarly by comparing $\textsc{inVal}$ and $\textsc{outVal}$.

\input{dbht2}

%% file: algo_dbht.tex
\begin{algorithm}[!t]
\DontPrintSemicolon
\fontsize{7pt}{7pt}\selectfont
\KwData{$n \times n$ dissimilarity matrix $D$, weighted undirected planar graph $G$ generated by TMFG, bubble tree $T$}
computeDirection(root of $T$)\; \label{alg:parDBHT:direction}
\revised{$v.g=(-\infty, -\infty), v.q=(-\infty, -\infty)$ $\forall$ vertices $v \in G$} \;\label{alg:parDBHT:initv}
\revised{\textsc{BB} = set containing $T$'s nodes}\;\label{alg:parDBHT:getbb}
\revised{\textsc{cvgBB} = \{$\bubble \in \textsc{BB}  : \bubble$.out-degree$=$0\} }\;\label{alg:parDBHT:getcvg}
\PFor{$\bubble \in$ \textsc{BB} }{\label{alg:parDBHT:bfsfor}
    \revised{BFS($T$, $\bubble$)}\;\label{alg:parDBHT:bfsfor2}
}
\tcp*[h]{all-pairs shortest paths}\; %
$A$ = allPairsShortestPaths($G$, $D$)\;\label{alg:parDBHT:apsp}
\tcp*[h]{assign vertices to converging bubbles}\;
\PFor{$\bubble \in \textsc{cvgBB}$ }{\label{alg:parDBHT:cluster1}
\PFor{ $v \in \bubble$ }{
    $\chi$ = computeChi($v,\bubble$)\;
    \textsc{WriteMax}($v.g$, ($\chi$,$\bubble$))\;\label{alg:parDBHT:writemax1}
}
}
\revised{$V^0_\bubble = \{v \in G  : v.g = \bubble\}$ $\forall \bubble$}\;\label{alg:parDBHT:v0}
\revised{$v.g=(\infty , \infty)$ if $v.g$ is $(-\infty, -\infty)$ $\forall v$} \;\label{alg:parDBHT:initv2}
\PFor{$\bubble \in \textsc{cvgBB}$}{ \label{alg:parDBHT:cluster2}
\PFor{$v \rightharpoonup \bubble$ \revised{and $v.g == (\infty , \infty)$}}{
    $\bar{L}$ = computeAverageShortestPath($v,V^0_\bubble$)\;
    \textsc{WriteMin}($v.g$, ($\bar{L}$, $\bubble$))\;\label{alg:parDBHT:writemin}
}
}
\tcp*[h]{assign vertices to bubbles}\;
\PFor{$\bubble \in \textsc{BB}$ }{ \label{alg:parDBHT:bubble}
    $\chi_{\text{total}}=0$\;
    \For{$v \in \bubble$}{
        $\chi_v$ = computeChi'($v$, $\bubble$); $\chi_{\text{total}} \mathrel{{+}{=}} \chi_v$\;
    }
    \For{$v \in \bubble$}{
        \textsc{WriteMax}($v.q$, ($\chi_v/\chi_{\text{total}}$, $\bubble$))\; \label{alg:parDBHT:chi2write}
    }
}
\tcp*[h]{build hierarchy using complete linkage}\;
$\{\mathcal{Z}_1, \dots, \mathcal{Z}_n\}$ \tcc*[h]{initialize dendrogram nodes}\;\label{alg:parDBHT:initdendro}
\PFor{$\bubble^c \in \textsc{cvgBB}$}{ \label{alg:parDBHT:intrabb-start}
    $\textsc{BB}_{b^c} = \{v.q : v \in \bubble^c \}$\;
    \PFor{$\bubble \in \textsc{BB}_{b^c}$}{
    Let $\mathcal{Z}_{(\bubble^c,\bubble)}$ = completeLinkage($\{\mathcal{Z}_v : v.q =\bubble \wedge v.g = \bubble^c \}$)\;\label{alg:parDBHT:intrabb}
    }
}
\PFor{$\bubble^c \in \textsc{cvgBB}$ \label{alg:parDBHT:interbb-start}}{
    Let $\mathcal{Z}_{\bubble^c}$ = completeLinkage($\{\mathcal{Z}_{(\bubble^{c'},*)} :  \bubble^{c'} = \bubble^c\}$)\;\label{alg:parDBHT:interbb}
}
$\mathcal{Z}$ = completeLinkage($\{\mathcal{Z}_{\bubble^c} : \bubble^c \in  \textsc{cvgBB} \}$)\;\label{alg:parDBHT:intercluster}
dendrogram = computeHeight($\mathcal{Z}$)\; \label{alg:parDBHT:postprocessing}
\Return dendrogram\; \label{alg:parDBHT:end}
\caption{Parallel DBHT for TMFG}\label{alg:parDBHT}
\end{algorithm}

%% file: dbht2.tex
\subsection{Assigning Vertices}
\iffull
\showfull{In this subsection, we first describe the assignment rules of the DBHT algorithm, and then describe our parallel algorithm for computing the assignment.}
\fi
A \defn{converging bubble} is a bubble with only incoming edges, and no outgoing edges. 
Intuitively, converging bubbles are the ``end points of a directional path that follows the strongest connections"~\cite{song2012}, so they are considered %
the center of local clusters. 
The first level of clustering
assigns each vertex to a unique converging bubble.
If a vertex is in at least one converging bubble, then it is assigned to the converging bubble with
the strongest attachment $\chi(v,b) = \frac{\sum_{u\in b} w(u,v)}{3(|b|-2)}$, where $3(|b|-2)$ %
is the number of edges in the bubble $b$.
For TMFG, all bubbles have 6 edges, and so we can simplify it to $\chi(v,b) = \sum_{u\in b} w(u,v)$.
For a vertex that is not in any converging bubble, it is assigned to
the converging bubble that has the minimum mean average shortest path distance:
$$\bar{L}(v,b) = mean\{l_D(u,v)\ |\  u\in V^0_b \wedge v \rightharpoonup b\}.$$
$v \rightharpoonup b$ means $v$ is in some bubble that can reach $b$
in the directed bubble tree, and $l_D(u,v)$ is the shortest path distance 
from $u$ to $v$ in the TMFG using the dissimilarity measure $D$ as the edge weights.
For all bubbles $b$, we let $V^0_b$ be the set of vertices in converging bubbles 
that have already been assigned to $b$ from computing $\chi$.
Let vertices assigned to the same converging bubble $b$ in the procedure above be a \defn{group}. 

After this initial partitioning of vertices, we investigate how each of these
groups is internally structured by performing a second level of clustering. 
This time we assign each vertex to a unique bubble, but not necessarily a converging bubble.
A vertex $v$ is assigned to the bubble $b$ that maximizes a different attachment score %
$$\chi'(v,b) = \frac{\sum_{u\in b} w(u,v)}{\sum_{u',v'\in b}  w(u',v')}.$$

The pseudocode for our parallel DBHT algorithm is in \cref{alg:parDBHT}.
On Lines~\ref{alg:parDBHT:direction}--\ref{alg:parDBHT:chi2write},
we show how the two assignments are computed. %
 We first compute the necessary auxiliary data used in the 
vertex assignment computation.
On Line~\ref{alg:parDBHT:direction}, we compute the directions of bubble tree edges
using \cref{alg:direction}. 
On Line~\ref{alg:parDBHT:initv}, we initialize the fields $g$ and $q$ for each vertex
to $(-\infty, -\infty)$ to prepare for the \textsc{WriteMax} operation on them.
$g$ is the group assignment and $q$ is the bubble assignment. 
On Line~\ref{alg:parDBHT:getbb}, we initialize the set $\textsc{BB}$ containing the bubble nodes.
On Line~\ref{alg:parDBHT:getcvg}, we obtain all of the
converging bubbles $\textsc{cvgBB}$ by using a parallel filter based on the out-degree of the tree nodes.
On Lines~\ref{alg:parDBHT:bfsfor}--\ref{alg:parDBHT:bfsfor2}, we 
run a BFS in the directed bubble tree for each bubble in parallel,
and record for vertices in the bubbles which converging bubbles they can 
reach. This helps us to do the $v \rightharpoonup b$ computation 
later.
On Line~\ref{alg:parDBHT:apsp}, we compute all-pairs shortest paths
in the TMFG by running Dijkstra's shortest path algorithm 
from each bubble node in parallel.

Now that we have obtained all necessary auxiliary data, we start to compute the assignments.
On Lines~\ref{alg:parDBHT:cluster1}--\ref{alg:parDBHT:writemin}, we compute the 
vertex assignments to converging bubbles, i.e., the groups. 
\revised{First, we compute the group assignment for vertices in at least 
one converging bubble by computing all of the attachment scores $\chi)$ in parallel and using concurrent \textsc{WriteMax}es
to write the group assignment (Lines~\ref{alg:parDBHT:cluster1}--\ref{alg:parDBHT:writemax1}). 
}
On Line~\ref{alg:parDBHT:v0}, we compute $V^0_b$ for all converging bubbles $b$ by first using a parallel integer sort to sort the vertices by group assignment, and then using a parallel filter to obtain the start and end indices of the groups in the sorted vertex array.
\iffull
\showfull{On Line~\ref{alg:parDBHT:initv2}, we set $v.g=(\infty, \infty)$
for all $v$ whose group $g$ has not been assigned yet to prepare for $\textsc{WriteMin}$ operations.
Then, in parallel we loop over all converging bubbles $b$
and all vertices $v$ such that $v \rightharpoonup b$ and $v$ has 
not been assigned to any converging bubbles yet. In the loop,
we compute the mean average shortest path distance
$\bar{L}$ and use a \textsc{WriteMin} to write the converging bubble
with the smallest $\bar{L}$ to each vertex's assignment.
Next, on Lines~\ref{alg:parDBHT:bubble}--\ref{alg:parDBHT:chi2write},
we compute the bubble assignment. This is similar to 
Lines~\ref{alg:parDBHT:cluster1}--\ref{alg:parDBHT:writemax1}, except that
we keep track of the total edge weights in bubbles using the variable $\chi_{\text{total}}$,
and re-weight the attachments with $\chi_{\text{total}}$ when looking
for the largest attachment bubble.\footnote{
In their paper~\cite{song2012}, Song et al.\ keep vertices assigned to
a converging bubble on Lines~\ref{alg:parDBHT:cluster1}--\ref{alg:parDBHT:writemax1}
and do not re-assign them to a different bubble. However, their implementation does, and we choose
to be consistent with their implementation.
}}
\fi

\myparagraph{\revised{Example 3}} We continue with our example
in \cref{fig:example}(c).
Here we only have a single converging bubble $b_2$
because this is the only bubble with no outgoing edges.
Therefore, all vertices are assigned to this converging 
bubble on Lines~\ref{alg:parDBHT:cluster1}--\ref{alg:parDBHT:writemin}.
Next, we look at what happens on Lines~\ref{alg:parDBHT:bubble}--\ref{alg:parDBHT:chi2write}.
Vertices $4$, $5$, and $6$ are only in a single bubble, and so
they are assigned to the bubble they are in. 
Vertices $0$, $1$, $2$, and $3$ are in multiple bubbles,
and so we compute the $\chi'$ score and assign each vertex 
to the bubble with the maximum $\chi'$.
For example, for vertex $3$, we will 
compute $\chi'(3,b_2)$, $\chi'(3, b_3)$,
and $\chi'(3, b_4)$. In our example, we find that 
$\chi'(3, b_3)$ is the largest, and so we
assign vertex $3$ to $b_3$.

\subsection{Complete Linkage}
Now we describe how to obtain the dendrogram from the vertex assignments.
At a high level, we build the complete-linkage hierarchy at three levels: intra-bubble, 
inter-bubble, and inter-group~\cite{song2012}. 
For the complete-linkage algorithm, the distance between two vertices $u$ and $v$ is their shortest path distance $l_D(u,v)$,
and the distance between two set of vertices $V_1$ and $V_2$ is 
$d(V_1,V_2) = \max\{l_D(u,v)\ |\ u \in V_1, v \in V_2\}$.
We use the parallel complete-linkage algorithm by Yu et al.~\cite{Yu2021}.

The steps for building the dendrogram are shown on Lines~\ref{alg:parDBHT:initdendro}--\ref{alg:parDBHT:end} of \cref{alg:parDBHT}.
On Line~\ref{alg:parDBHT:initdendro}, we initialize $n$ dendrogram nodes,
one for each vertex in the TMFG. We define a \defn{subgroup} to be the set of vertices in a group that belong to the same bubble.
On Lines~\ref{alg:parDBHT:intrabb-start}--\ref{alg:parDBHT:intrabb}, for each subgroup we run the
complete-linkage algorithm on the dendrogram nodes corresponding
to vertices within the subgroup.
The result is a dendrogram
 for each subgroup. 
We use subgroups so that vertices within the same group, but in different bubbles will not be processed together, which is consistent with the sequential algorithm~\cite{song2012}.
Then, on Line~\ref{alg:parDBHT:interbb}, we run the
complete-linkage algorithm on dendrogram nodes in each group. 
The result is a dendrogram for each group.
Finally, on Line~\ref{alg:parDBHT:intercluster}, we run the
complete-linkage algorithm on the group
dendrogram nodes to obtain the final dendrogram.

\myparagraph{Dendrogram Heights}
In conventional complete-linkage clustering, the dendrogram height is the 
distance between the two merged clusters. 
The dendrogram's height requires that nodes 
closer to the dendrogram root
(clusters merged later) have a height at least the height of 
nodes further away from the root. 
However, since
we are concatenating three levels of dendrograms, 
the shortest path distance might not satisfy the
height requirement. 
For example, the maximum shortest path distance %
might be larger between two vertices in a bubble than between two 
bubbles. 
As a result, we will have to re-assign
heights to the dendrogram nodes. We use the same height
assignment in the implementation by Aste~\cite{dbhtcode}, which ensures that 
all nodes that contain exactly one group are at the same height.

For inter-group dendrogram nodes merged on Line~\ref{alg:parDBHT:intercluster}, the height
is the number of converging bubbles in its descendants. 
This can be computed in a top-down traversal from the root $\mathcal{Z}$. 
Intra-bubble and inter-bubble dendrogram nodes that belong to the same converging bubble $b^c$ have heights chosen from
$[\frac{1}{n_b-1},\frac{1}{n_b-2},\dots,\frac{1}{2},1]$,
where $n_b$ is the number of vertices assigned to converging bubble $b^c$. 
Each bubble's dendrogram nodes have a contiguous segment of the heights in the height list. 
To obtain the dendrogram height for descendants of each $\mathcal{Z}_{(b^c)}$,
we sort these dendrogram nodes such that nodes merged on Line~\ref{alg:parDBHT:intrabb}
appear before nodes merged on Line~\ref{alg:parDBHT:interbb}. Nodes merged on Line~\ref{alg:parDBHT:interbb}
are sorted by the distance when they are merged. Nodes merged on Line~\ref{alg:parDBHT:intrabb}
are sorted first by bubble assignment and then by the distance when they are merged.
Now all dendrogram nodes in the same bubble and converging bubble are contiguous. 
We then assign heights  $[\frac{1}{n_b-1},\frac{1}{n_b-2},\dots,\frac{1}{2},1]$ 
to the dendrogram nodes in the sorted order. 
This sorted order ensures that the inter-group hierarchy is below the inter-bubble hierarchy in the dendrogram.

\myparagraph{\revised{Example 4}} 
In \cref{fig:example}(c),
since $\{2,4\}$, $\{0,3,6\}$,  and $\{1,5\}$
are all assigned to the same bubble, we first have
three dendrograms $\mathcal{Z}_{(b_2, b_1)}$, $\mathcal{Z}_{(b_2, b_3)}$, and $\mathcal{Z}_{(b_2, b_4)}$. 
Then we get a single dendrogram $\mathcal{Z}_{b_2}$
by running complete linkage on the three dendrogram roots.
In this example, we only have a single converging bubble, and so we are done. If there were multiple converging bubbles, our algorithm
would run complete linkage on the group dendrogram roots, which would all be at the same height 1, to produce the final dendrogram.
The resulting dendrogram is shown in \cref{fig:example}(d).

%% file: analysis.tex
\section{Analysis}
In this section, we  analyze the complexity of our parallel algorithms using the work-span model defined in Section~\ref{sec:background}.

We use $\rho$ to denote the number of rounds required by the parallel TMFG algorithm, i.e., the number of times we  iterate the while loop on Line~\ref{alg:partmfg:while} of \cref{alg:partmfg}. 
We use $\pi$ to denote the size of the $\prefix$.

\myparagraph{Parallel TMFG}
First, we analyze the complexity of the parallel TMFG algorithm given in \cref{sec:partmfg}. The first step of computing the initial four vertices takes $O(n^2)$ work and $O(1)$ span. Initializing the \textsc{Gains} array takes $O(n)$ work and $O(1)$ span because initially there are $n-4$ nodes in $V$ and four faces in $\faces$.

Now we analyze the while loop.
Line~\ref{alg:partmfg:getprefix} takes $O(n \log n)$ work and $O(\log n)$ span for parallel sorting. 
Lines~\ref{alg:partmfg:conflict}--\ref{alg:partmfg:restv2} takes $O(\pi \log \pi)$ time and $O(\log \pi)$ span for parallel filtering and sorting.
In the for-loop, each vertex in $L$ contributes $O(1)$ work and span on Lines~\ref{alg:partmfg:13}--~\ref{alg:partmfg:14}.
On Lines~\ref{alg:partmfg:updategain}--\ref{alg:partmfg:updategainloop}, we maintain a sorted list of vertices for each face. Every time we create a new face, we take $O(n \log n)$ work to sort all vertices by their gains with respect to this face, and we maintain a linked list of sorted vertices for each face. Each vertex stores pointers to its positions in these linked lists.  %
When we insert a vertex to a face, we remove the vertex from the linked lists of all other faces. Since there are $O(n)$ faces, the total work of sorting is $O(n^2 \log n)$ and the total work of removing vertices from linked lists is $O(n^2)$ because each vertex is removed exactly once for each face.
The span of this algorithm is $O(\rho\log n)$ because the span of each round is $O(\log n)$.

\myparagraph{Bubble Tree Construction}
The bubble tree is constructed by the highlighted lines in \cref{alg:partmfg} (which calls \cref{alg:updatebubbletree}).
Building the bubble tree takes $O(n)$ work because we insert $n$ vertices, and
each insertion takes $O(1)$ work. 
Since we update the bubble tree on each round in $O(1)$ span, the total span is $O(\rho)$. 

\myparagraph{Directing Edges}
The directions of bubble tree edges are computed in \cref{alg:direction}.
Computing the bubble tree direction takes $\Theta(n)$ work because for each bubble, we perform $O(1)$ work (and span), and there are $n-3$ bubbles in total. The span of computing the directions is asymptotically the same as the height of the bubble tree. This is because on each round of \cref{alg:partmfg} we only add tree edges adjacent to existing bubble node, so the height can increase by at most 2 in each round, and thus the span of this step is $O(\rho)$.

\myparagraph{Assigning Vertices}
Now, we give the complexity of the vertex assignment step in parallel DBHT algorithm (Lines~\ref{alg:parDBHT:getcvg}--\ref{alg:parDBHT:chi2write} in \cref{alg:parDBHT}). 
Lines~\ref{alg:parDBHT:getcvg}, \ref{alg:parDBHT:cluster1}--\ref{alg:parDBHT:initv2}, and \ref{alg:parDBHT:writemin}--\ref{alg:parDBHT:chi2write}
take $O(n)$ work because there are $O(n)$ bubbles, and each bubble has four vertices. %
Lines~\ref{alg:parDBHT:cluster2}--\ref{alg:parDBHT:writemin} takes $O(n^2)$ work because there are at most $O(n)$ converging bubbles, and each vertex can reach $O(n)$ converging bubbles in the worst case.
 Lines~\ref{alg:parDBHT:cluster1}--\ref{alg:parDBHT:chi2write} have span $O(1)$ except Line~\ref{alg:parDBHT:v0}  which has $O(\log n)$ span for integer sorting. 
 Lines~\ref{alg:parDBHT:bfsfor}--\ref{alg:parDBHT:bfsfor2} take $O(n^2)$ work because we 
run $O(n)$ BFS's. The span of running a BFS is $O(\rho\log n)$ because the bubble tree has diameter $O(\rho)$~\cite{BlellochM97}.

The all-pairs shortest paths (APSP) on Line~\ref{alg:parDBHT:apsp} can either be computed by running single-source shortest path (SSSP) from all nodes or can be computed by a dedicated APSP algorithm.
The APSP and SSSP problems, both in the sequential and parallel settings, have been well-studied~\cite{meyer2003delta, dong2021shortest, dhulipala2017julienne, Klein1997,Cohen2000,Sinha,Seidel,PR94,Madduri07, doi:10.1137/1.9781611976465.17}. 
There are also APSP and SSSP algorithms specifically  designed for planar graphs, which have better complexity than their general counterparts~\cite{HENZINGER19973, pantziou1990efficient, Federicksonshortestplanar, traff1996simple, 366861shortest, chen2000shortest, djidjev1991computing,djidjev1996efficient, fakcharoenphol2006planar,klein1994faster, 10.1145/1721837.1721846, 10.1007/978-3-642-15781-3_18, doi:10.1137/1.9781611975031.34}. 
While Federickson~\cite{Federicksonshortestplanar} gives a sequential algorithm that computes APSP in $O(n^2)$, we are not aware of any parallel algorithms that achieve the same work with lower depth.
Some examples for APSP on planar undirected graphs include an algorithm by Pan and Reif that runs in $O(n^2 \log n)$ work and $O(\log^2 n)$ span~\cite{pan1989fast}, and an algorithm by Henzinger et al.\ that runs in
 $O(n^{7/3} \log n \log(nD))$ work and
 $O(n^{2/3} \log^{7/3} n \log(nD))$ span, where $D$ is the sum of the absolute values of the edge weights~\cite{HENZINGER19973}. 
Let the work of the APSP algorithm used be $W_{APSP}$ and the span be $S_{APSP}$.

\myparagraph{Complete Linkage}
Finally, we consider the complete linkage step. Let $\rho_l$ be the number of rounds used by the parallel complete linkage algorithm~\cite{Yu2021}. The span is $O(\rho_l\log n)$ and the work is $O(\rho_l n^2)$ in the worst case.  Yu et al.~\cite{Yu2021} show that in practice, the running time of the complete linkage algorithm is close to $O(n^2)$. 

\myparagraph{Summary}
In total, TMFG takes $O(n^2 \log n)$ work and $O(\rho \log n)$ span  and DBHT takes $O(W_{APSP} + \rho_l n^2)$ work and $O(S_{APSP} + \rho+\rho_l\log n)$ span. 
Compared to Song et al.~\cite{song2012}, our algorithm has a lower span in all steps and lower work in bubble tree construction and direction computations. Song et al.\ take $O(n^2)$ work for  bubble tree construction and direction computations, while our algorithm takes $O(n)$ work. We show in \cref{sec:exp} that all of our steps are faster. The fastest part of their DBHT code was the APSP computation, and all other parts were up to orders of magnitude slower; we have significantly reduced the running times of all parts of the DBHT code, and now our bottleneck is APSP.

%% file: exp.tex
\section{Experiments}\label{sec:exp}

\myparagraph{Testing Environment} We perform experiments on
a \texttt{c5.24xlarge} machine on Amazon EC2, with $2$ Intel Xeon Platinum 8275CL (3.00GHz) CPUs for a total of 48
hyper-threaded cores, and 192 GB of RAM.  By default, we use all cores
with hyper-threading.  For the C++ code tested, we use the \texttt{g++} compiler (version 7.5)
with the \texttt{-O3} flag, and use ParlayLib~\cite{blelloch2020parlaylib} 
for parallelism in our code. 

\begin{itemize}[topsep=1pt,itemsep=0pt,parsep=0pt,leftmargin=10pt]
  \item \defn{\textsc{pmfg-dbht}} is the existing sequential PMFG and DBHT implementation in MATLAB, which we obtained online~\cite{dbhtcode}. The bottleneck of \textsc{pmfg-dbht} is in constructing 
the PMFG. \revised{We tried implementing the same PMFG construction
algorithm in C++, but found it slower than the MATLAB 
implementation, and so we report the MATLAB runtime.}
\item \defn{\textsc{seq-tdbht}} is the state-of-art implementation of sequential TMFG and DBHT in MATLAB, which we obtained online~\cite{dbhtcode}. 
    \item \defn{\textsc{par-tdbht}} is our implementation of parallel TMFG and DBHT in C++. We tested prefixes of size 1, 2, 5, 10, 30, 50, and 200.
    \item \defn{\textsc{comp}} and \defn{\textsc{avg}} are the parallel C++ complete and average linkage implementations, respectively, by Yu et al.~\cite{Yu2021}.
    \item \defn{\textsc{k-means}} \revised{is the scalable $k$-means++~\cite{bahmani2012scalable} implementation~\cite{kmeansgit} in C, parallelized using OpenMPI. }
    \item \defn{\textsc{k-means-s}} is the parallel $k$-means++ implementation from scikit-learn with a preprocessing step that computes a spectral embedding, which constructs its affinity matrix using a nearest-neighbors graph~\cite{lucinska2012spectral}. We choose the number of nearest neighbors that gives the best clustering quality. 
    If the true number of clusters is $c$, then we project the original data onto the $c$-dimensional space, which we find to be a good heuristic for finding good clusters. \revised{We also tried a C++ implementation using Eigen~\cite{guennebaud2010eigen}, but found the scikit-learn version to be faster.}
\end{itemize}

\myparagraph{Evaluation}
We evaluate the clustering quality using the Adjusted Rand Index (ARI)~\cite{Hubert1985} and Adjusted Mutual Information (AMI)~\cite{Vinh2010} scores. In our experiments, AMI showed similar trends as ARI, and so we only show the plots for ARI. 
Let $n_{ij}$ be the number of objects in the ground truth cluster $i$ and the cluster generated by the algorithm $j$, $n_{i*}$ be $\sum_j n_{ij}$, $n_{*j}$ be $\sum_i n_{ij}$, %
and $n$ be $\sum_i n_{i*}$. The ARI is computed as 
$$\frac{\sum_{i,j} {n_{ij}\choose 2} - [\sum_i {n_{i*}\choose 2} \sum_j {n_{*j}\choose 2} ] / {n\choose 2}}{\frac{1}{2}[\sum_i {n_{i*}\choose 2} +\sum_j {n_{*j}\choose 2} ] - [\sum_i {n_{i*}\choose 2} \sum_j {n_{*j}\choose 2} ]  /   {n \choose 2}}.$$
The ARI score is $1$ for a perfect match, and its expected value is $0$ for random assignments.

When computing the ARI for the hierarchical methods, we cut the
dendrogram such that the number of resulting clusters is the same as the
number of ground truth clusters. For the $k$-means methods, we set $k$ to be equal to the number of ground truth
clusters.

\myparagraph{Data sets} We show results on 18  data sets from the UCR Time Series Classification Archive~\cite{UCRArchive2018}, including the three largest data sets Crop, ElectricDevices, and StarLightCurves (we removed duplicate points from these three data sets). 
The data sets are summarized in \cref{tab:datasets}. 
Since the UCR Archive is designed for classification tasks, not all data sets are suitable for clustering. 
We choose the two largest data sets as well as 16 data sets which have ARI scores of at least $0.2$ for \textsc{k-means}.

We also collected the closing daily prices of 1614 US stocks between Jan.\ 1, 2013 and Jan.\ 1, 2019 (1761 trading days) using the Yahoo Finance API. We used the Industry Classification Benchmark (ICB) to obtain ground truth clusters.

We used the Pearson correlation coefficient $p$ for the similarity measure and $d = \sqrt{2(1-p)}$ for the dissimilarity measure~\cite{marti2021review}. For normalized and zero-mean vectors, $d$ is the same as the squared Euclidean distance.

\begin{table}
\footnotesize
  \caption{Summary of UCR data sets used in the experiments. $n$ is the number of objects, and $L$ is the length or size of the object.}
  \label{tab:datasets}
\begin{tabular}{ccccc}
\toprule
ID & Name & $n$ & $L$ & \# of classes \\ \hline \rowcolor{Gray}
1 & Mallat & 2400 & 1024 & 8 \\
2 & UWaveGestureLibraryAll & 4478 & 945 & 8 \\\rowcolor{Gray}
3 & NonInvasiveFetalECGThorax2 & 3765 & 750 & 42 \\
4 & MixedShapesRegularTrain & 2925 & 1024 & 5 \\\rowcolor{Gray}
5 & MixedShapesSmallTrain & 2525 & 1024 & 5 \\
6 & ECG5000 & 5000 & 140 & 5 \\\rowcolor{Gray}
7 & NonInvasiveFetalECGThorax1 & 3765 & 750 & 42 \\
8 & StarLightCurves & 9236 & 84 & 2 \\\rowcolor{Gray}
9 & HandOutlines & 1370 & 2709 & 2 \\
10 & UWaveGestureLibraryX & 4478 & 315 & 8 \\\rowcolor{Gray}
11 & CBF & 930 & 128 & 3 \\
12 & InsectWingbeatSound & 2200 & 256 & 11 \\\rowcolor{Gray}
13 & UWaveGestureLibraryY & 4478 & 315 & 8 \\
14 & ShapesAll & 1200 & 512 & 60 \\\rowcolor{Gray}
15 & SonyAIBORobotSurface2 & 980 & 65 & 2 \\
16 & FreezerSmallTrain & 2878 & 301 & 2 \\\rowcolor{Gray}
17 & Crop & 19412  & 46 &  24\\
18 & ElectricDevices & 16160 & 96 & 7 \\
\bottomrule
\end{tabular}
\end{table}

\begin{figure}
    \centering
    \vspace{-5pt}
    \includegraphics[width = 0.8\columnwidth]{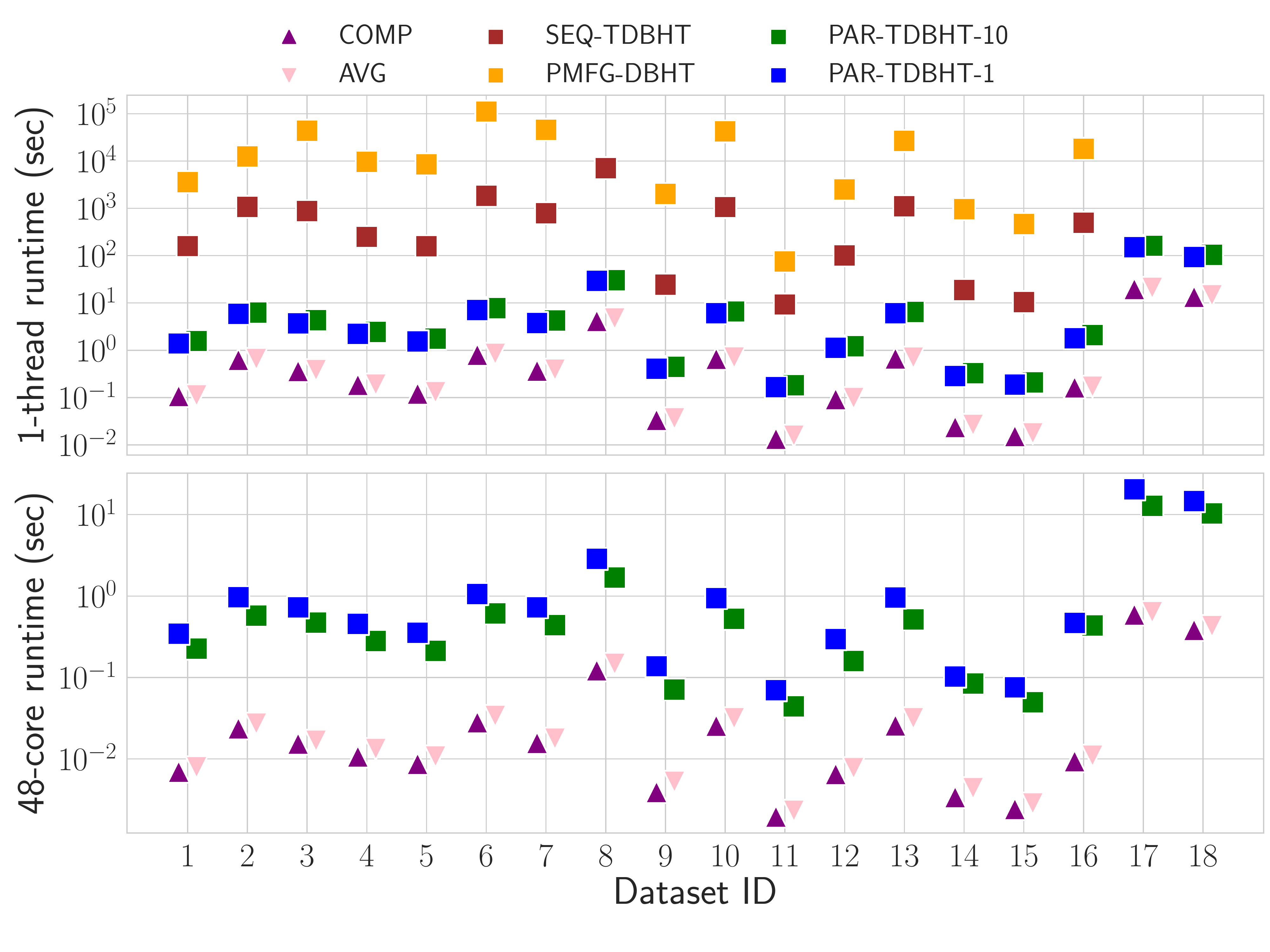}
    \caption{Time (seconds) of different methods on UCR data sets. The top plot shows run times on a single thread and the bottom plot shows run times on 48 cores with hyper-threading. \textsc{PMFG-DBHT} and \textsc{SEQ-TDBHT} timed out for data sets 17 and 18. \textsc{PMFG-DBHT} also timed out for data set 8.}%
    \label{fig:time}
    \vspace{-5pt}
\end{figure}

\begin{figure}[t]
\vspace{-4pt}
    \centering
    \includegraphics[width = 0.8\columnwidth]{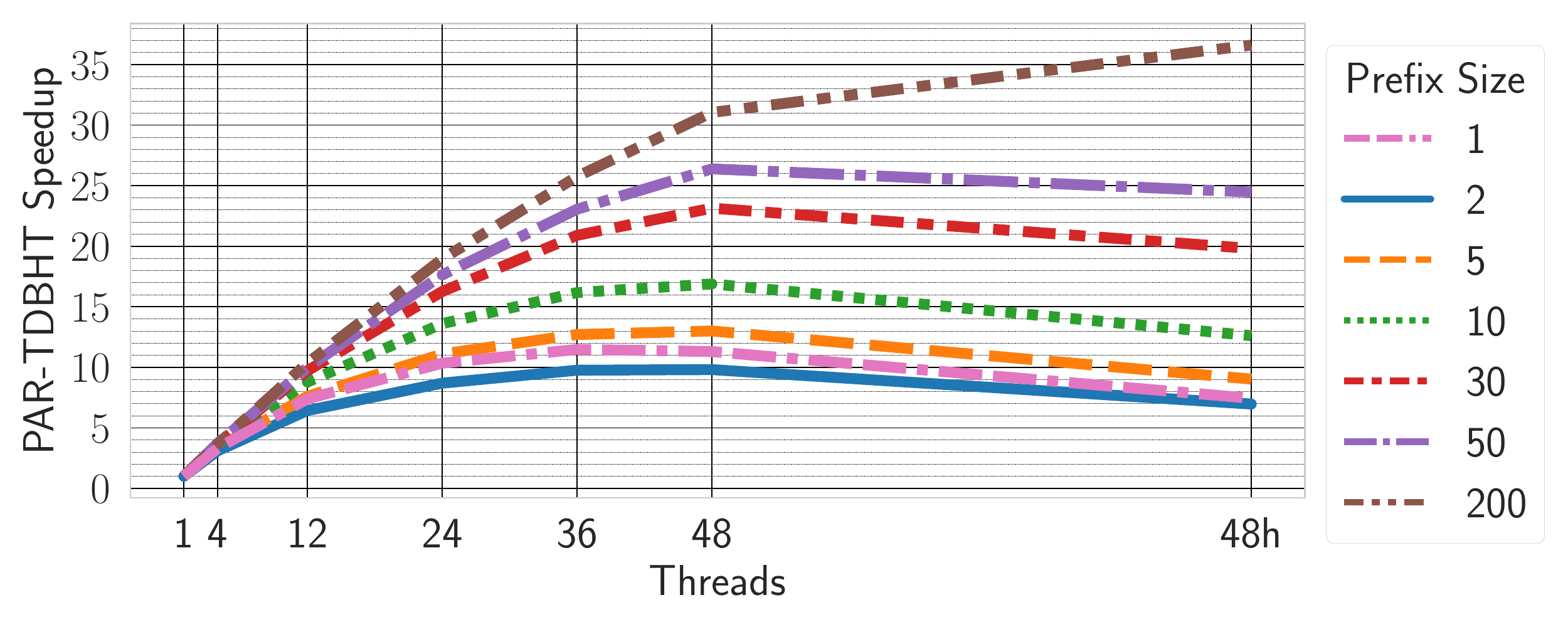}
    \caption{Self-relative parallel speedup vs.\ thread counts for \textsc{par-tdbht} with different prefix sizes on the Crop data set.
"48h" indicates 48 cores with two-way hyper-threading. The speedups of some data sets decrease when hyper-threading because there is not enough work and the overhead of hyper-threading is high relative to the work of the algorithm.} 
    \label{fig:speedup}
\end{figure}

\begin{figure}[!t]
    \centering
    \vspace{-5pt}
    \includegraphics[width = 0.8\columnwidth]{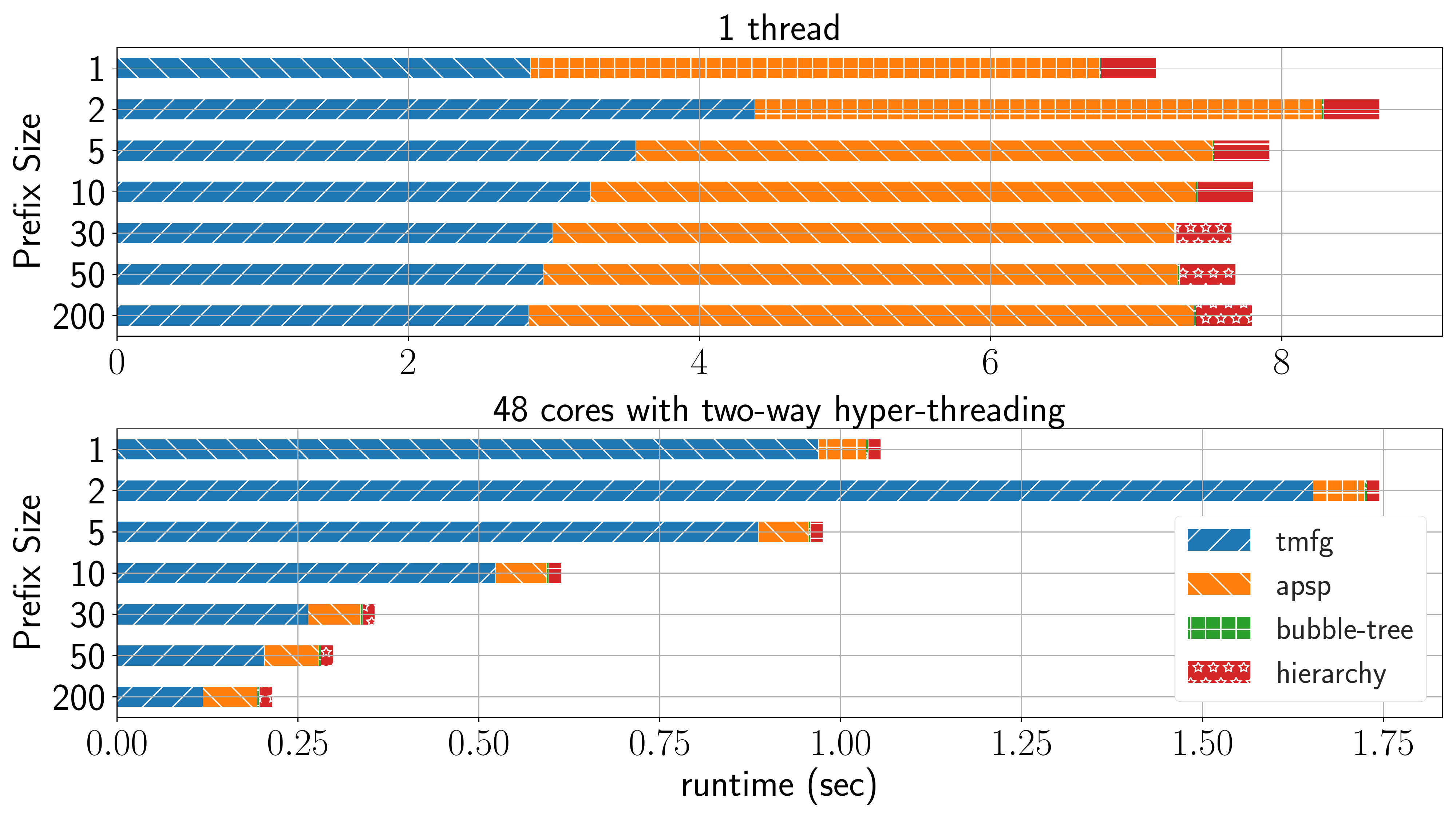}
    \caption{Breakdown of runtime across different steps of
our algorithm on ECG5000. 
"tmfg" corresponds to TMFG construction (\cref{alg:partmfg});
"apsp" corresponds to the all-pairs shortest paths computation;
"bubble-tree"
corresponds to computing the direction of bubble tree edges and 
assigning vertices to bubbles; and
"hierarchy"
corresponds to running the complete-linkage subroutine.
}
    \label{fig:decomposition}
\end{figure}

\subsection{Runtime}
We show the running times of all hierarchical clustering algorithms and data sets in \cref{fig:time} 
(sequential times are in the top plot and parallel times are in the bottom plot). 
\textsc{par-tdbht-1} is \textsc{par-tdbht} with a prefix of size 1 and \textsc{par-tdbht-10}
is \textsc{par-tdbht} with a prefix of size 10 (we chose this prefix size for most experiments as it gives a good tradeoff between speed and cluster quality).
Since \textsc{pmfg-dbht} and \textsc{seq-tdbht} are sequential, we only include it in the top plot.

We see that \textsc{pmfg-dbht} and \textsc{seq-tdbht} are both orders of magnitude slower than all other 
methods. 
\textsc{pmfg-dbht} is 458--15586x slower than \textsc{par-tdbht-1} 
and 414--14254x slower than \textsc{par-tdbht-10} on a single thread.
\textsc{seq-tdbht} is 56--276x slower than \textsc{par-tdbht-1}
and 50--235x slower than \textsc{par-tdbht-10} on a single thread. 
On 48 cores with hyper-threading, 
\textsc{seq-tdbht} is 136--2483x slower than \textsc{par-tdbht-1}
and 226--4487x slower than \textsc{par-tdbht-10}.
We discuss the reasoning for this speedup in the Runtime Decomposition section below. 
\textsc{par-tdbht-1} and \textsc{par-tdbht-10}
 are slower than \textsc{avg} and \textsc{comp},
 which is expected because DBHT uses complete-linkage clustering
 as a subroutine. However, we see in \cref{sec:compare}
 that \textsc{par-tdbht-1} and \textsc{par-tdbht-10} give
 significantly better clusters than \textsc{avg} and \textsc{comp}
 on most data sets.
 
\textsc{k-means-s} and \textsc{k-means} are not included because they do not generate a dendrogram, and 
one would need to run the algorithm multiple times to obtain clusters of different scales.
\revised{On average across the data sets, one run of \textsc{k-means} is 1.82x faster than \textsc{par-tdbht-1} 
and 1.31x faster than \textsc{par-tdbht-10} 
on 48 cores with hyper-threading; and 
one run of \textsc{k-means-s} is 10.33x slower than \textsc{par-tdbht-1} 
and 16.23x slower than \textsc{par-tdbht-10}.
We used the 12-thread times for \textsc{k-means-s} because it
became slower past 12 threads due its parallel overheads.
Though we are slower than \textsc{k-means}, \textsc{k-means} is not hierarchical
and also not deterministic.}

\myparagraph{Scalability with Thread Count} 
In \cref{fig:speedup}, we show the scalability of our algorithm
vs.\ thread count on the largest Crop data set. 
\textsc{par-tdbht} with a prefix size of 200 achieves a self-relative speedup of 36.6x
on 48 cores with two-way hyper-threading.
In general, a larger
prefix size results in higher scalability, because more insertions
in the TMFG construction can be processed in parallel. 
However, using a prefix of size 2 is actually slower than using a prefix of size 1,
which corresponds to the exact TMFG, for the following reason.
When we only have a prefix of size 1, our implementation
only needs to find the best vertex-face pair to insert using a parallel maximum, but
when the prefix size is larger than 1, the algorithm needs to first
sort all vertex-face pairs, and then find a prefix of
the sorted pairs to insert. A prefix of size 2 is 
small, and so there is not enough additional parallelism to 
offset the overheads of sorting.
In our experiments, using a prefix of size 5 or larger gives similar or better runtimes than using exact TMFG.
Our largest self-relative speedup is on StarLightCurves (data set 8), where \textsc{par-tdbht} with a prefix size of 200 achieves a speedup of 41.57x.

\myparagraph{Scalability with Data Size}
We observe that on our data sets, the
\textsc{par-tdbht} runtimes scale with the data size $n$ approximately as a function of $O(n^{2.22})$ on a single thread and $O(n^{1.79})$
on 48 cores with two-way hyper-threading. 
The scaling for parallel runtimes is better than for the single-threaded runtimes because we get more parallel speedups for larger values of $n$. 

\myparagraph{Runtime Decomposition}
We
show the breakdown of runtime across different steps of
our algorithm in \cref{fig:decomposition} on the ECG5000 data set 
(the different steps are described in the figure caption). 
Sequentially, the majority of the runtime is in the "tmfg" and "apsp" steps, while
in parallel the majority of the runtime is in the "tmfg" step, especially for small 
prefix sizes where TMFG construction has limited parallelism. 
When the prefix size is larger, the runtime of the "tmfg" step is significantly shorter.
Our "bubble-tree" for TMFG is very fast and the runtime is too small to be visible
in the plot.

For comparison, \textsc{seq-tdbht} requires 628s for "tmfg", 9s for "apsp", 69s for "bubble-tree", and 1136s for "hierarchy"
on the same data set, and thus all steps of \textsc{par-tdbht}, even on one thread, are faster than \textsc{seq-tdbht}.
Our "tmfg" step is faster because we use optimized data structures to update the gain table that do not require looping over all faces. 
\revised{The "apsp" step is faster because we use a different graph data structure. \textsc{seq-tdbht} uses Johnson's algorithm
from the Boost Graph Library, while
we used the faster Dijkstra's algorithm from the same library. 
We also tried Dijkstra's algorithm for SEQ-TDBHT and found that in MATLAB's Boost Graph Library,
Dijkstra's algorithm is on average 16\% slower than Johnson's algorithm (possibly due to having to convert the MATLAB graph to C++ format many more times).}
Our "bubble-tree" is step is faster because we optimized
it for the TMFG topology. While this step is much slower
than "apsp" in \textsc{seq-tdbht}, its time becomes negligible
in \textsc{par-tdbht}.
Our "hierarchy" step is faster because we use the optimized
complete-linkage algorithm by Yu et al.~\cite{Yu2021}, whereas \textsc{seq-tdbht} uses a less efficient implementation.
Across all data sets, our "tmfg" step is 
15.62--9629x faster than \textsc{seq-tdbht}
and the remaining steps are together 1278--14049x faster than \textsc{seq-tdbht} on 48 cores with hyper-threading.
On a single thread,  our "tmfg" step is 
12.19--319.56x faster than \textsc{seq-tdbht}
and the remaining steps are together 59.42--431.17x faster than \textsc{seq-tdbht}.
In \textsc{par-tdbht}, TMFG and APSP take most of the execution time when running sequentially, which means that the running time could potentially be improved by using a more sophisticated APSP implementation. However, in the baseline implementation~\cite{song2012}, APSP takes only a small fraction of running time, and the bottlenecks are in other steps. This means that our algorithm has reduced the bottleneck of the original algorithm.

\begin{figure}[!t]
    \centering
    \vspace{-5pt}
    \includegraphics[width = 0.9\columnwidth]{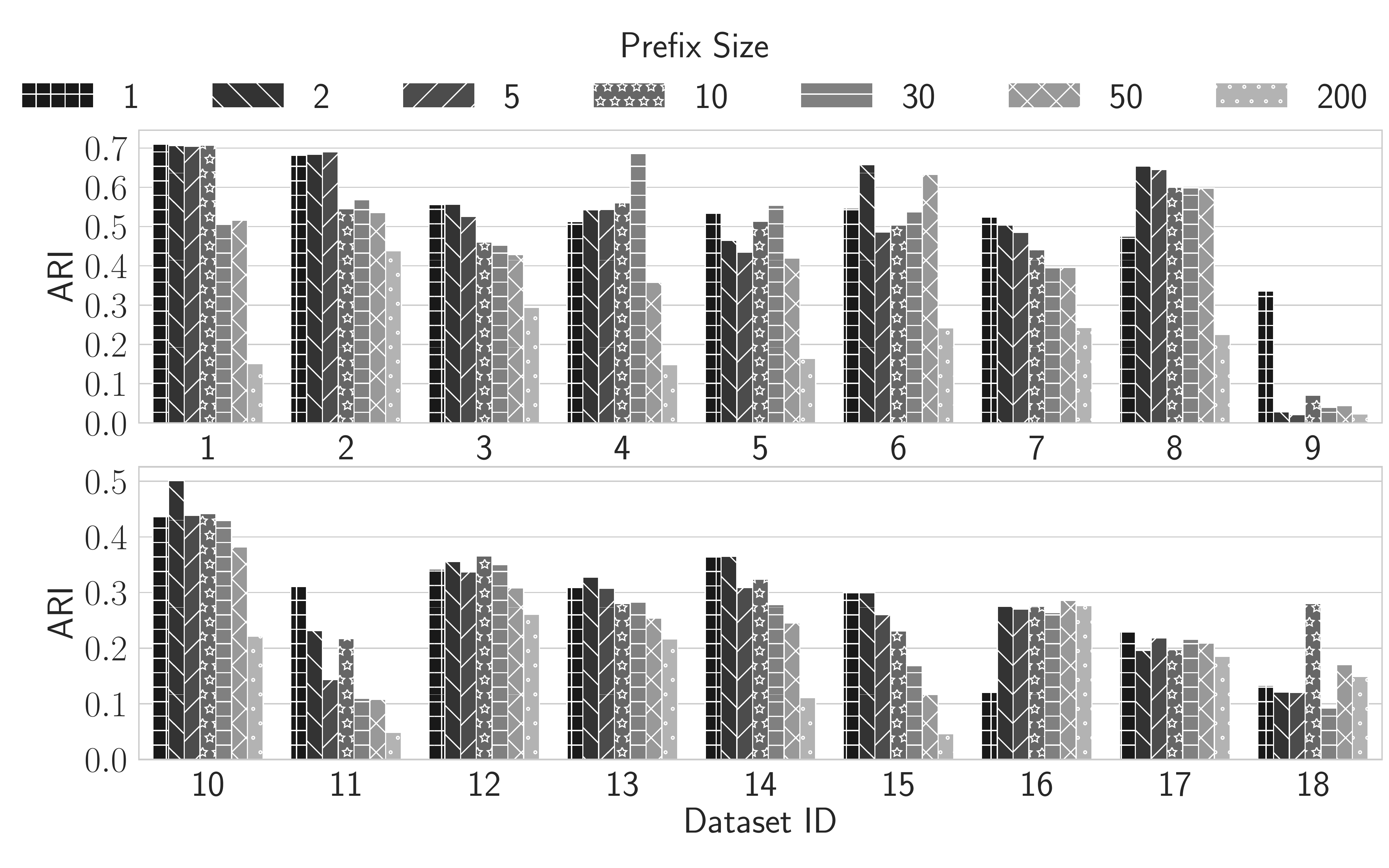}
    \caption{Clustering quality (ARI score) of \textsc{par-tdbht}. Different shades represent different prefix sizes. 
    }
    \label{fig:prefix}
\end{figure}

\subsection{Clustering Quality}\label{sec:compare}
\myparagraph{Prefix Size and Clustering Quality}  
\revised{Our prefix-based
TMFG algorithm is able to produce filtered graphs with weight very close to, and
sometimes even higher than, the sequential TMFG and PMFG algorithms.
In our experiments, our prefix-based TMFG algorithm produces graphs with edge weight sums that are 92.1--100.3\% of the edge weight sums produced by PMFG. If we only consider prefix sizes up to 50, then the edge weight sums are 97.1--100.3\% of the edge weight sums produced by PMFG.}
\iffull
\showfull{We show a plot of edge weight sums for different prefix sizes in Figure~\ref{fig:edge-sum}.}
\begin{figure}
    \centering
    \includegraphics[width=\columnwidth]{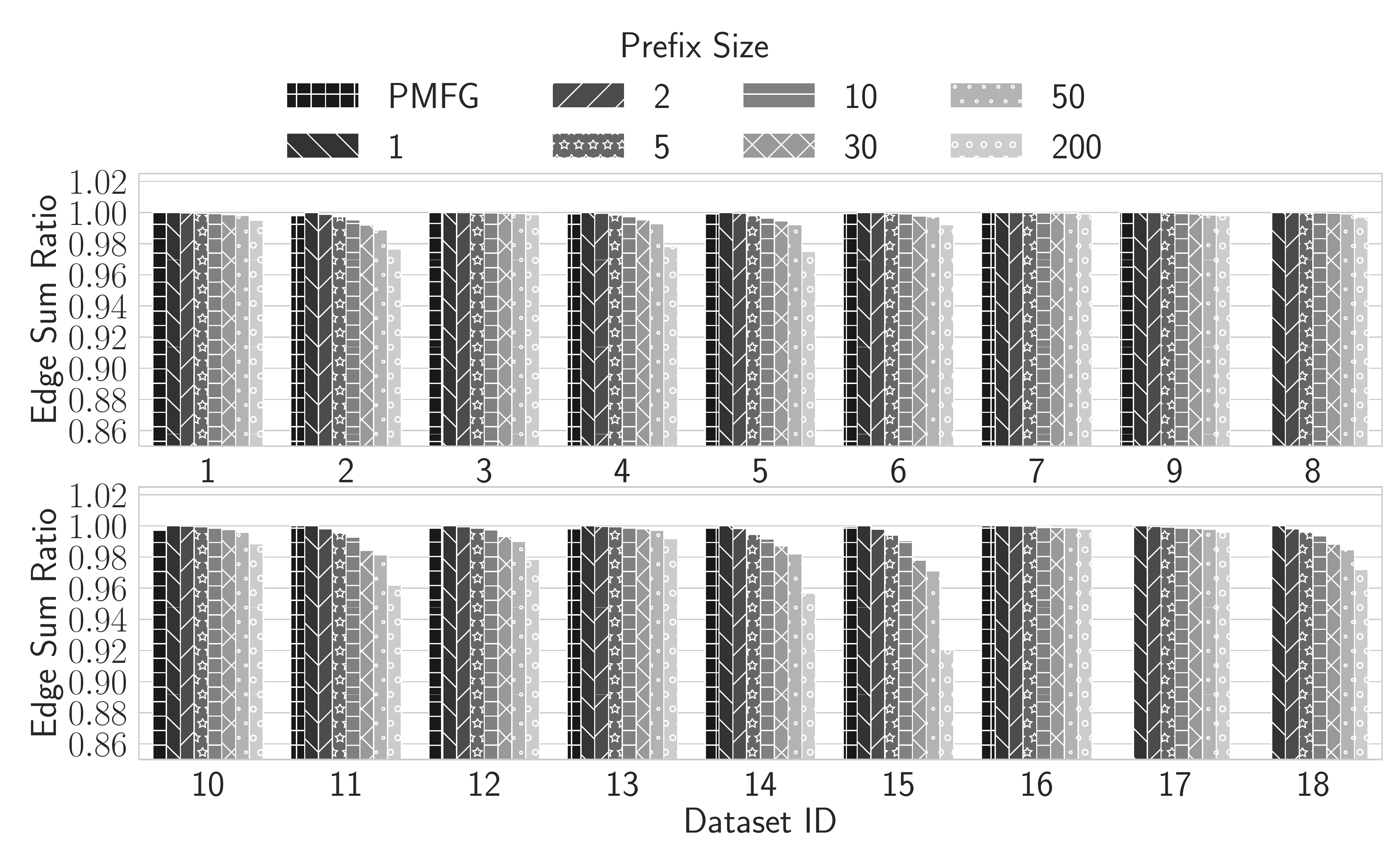}
    \caption{The ratio of edge sum compared to the edge sum of \textsc{SEQ-TMFG}. }
    \label{fig:edge-sum}
\end{figure}
\fi

We present the ARI of using varying prefix sizes in \cref{fig:prefix}.
We find that \textsc{par-tdbht}
with a prefix of size greater than 1 gives similar, and sometimes even better ARI than using a prefix of size 1 (which correponds to using the exact TMFG). \revised{Generally, using a larger prefix size results 
in lower ARI, but sometimes a larger prefix size can 
result in better quality as the clusters could become less sensitive to noise. We provide a concrete example of this behavior in the Appendix.}
For smaller data sets (e.g., data sets 9, 11, and 15), the ARI degradation is larger.
This is because the prefix is a larger percentage of all edges in the filtered graph.
For larger data sets (e.g., data sets 2, 6, 8, 10, 13, 17, and 18), the ARI degradation is smaller.

\begin{figure}[!t]
    \centering
    \vspace{-5pt}
    \includegraphics[width = 0.9\columnwidth]{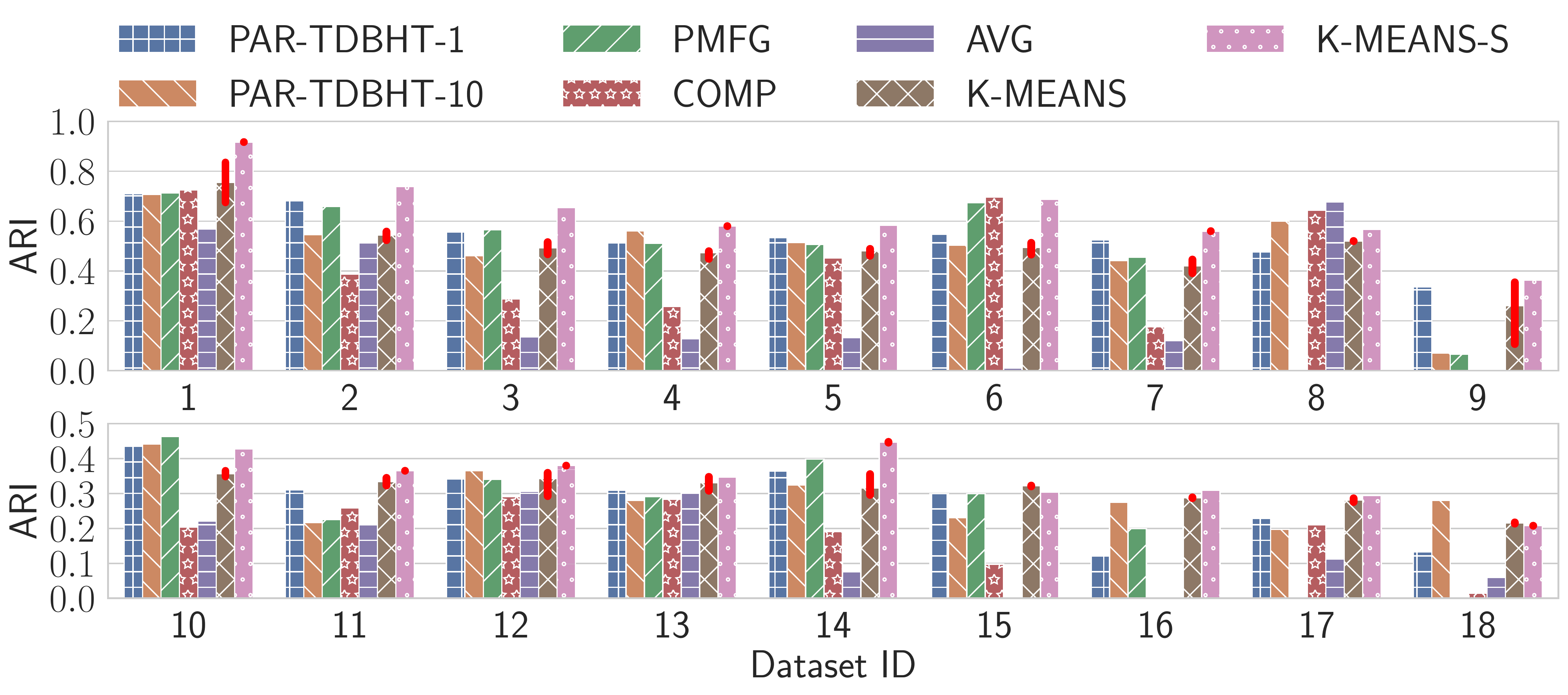}
    \caption{Clustering quality of different methods on UCR data sets. A few bars for \textsc{comp} and \textsc{avg} are hard to observe because their ARIs are close to 0. 
    \textsc{PMFG} timed out for data sets 8, 17, and 18. \revised{The error bar on \textsc{k-means} shows the range of ARI scores obtained when running it with different numbers of threads.}
    } %
    \label{fig:quality}
\end{figure}

\myparagraph{Hierarchical Methods} 
We show the clustering quality of all of the methods and data sets in \cref{fig:quality}. %
 We see that \textsc{par-tdbht-1} and \textsc{par-tdbht-10} often generate higher-quality clusters than both of the other hierarchical clustering algorithms, \textsc{comp} and \textsc{avg}.
On data sets where the number of ground truth clusters is very small, such as data sets  9, 15, and 16,
\textsc{comp} and \textsc{avg} have low ARI scores. 
This is because \textsc{comp} and \textsc{avg} are sensitive to agglomeration decisions, which only use local information, and when these decisions are wrong, they lead to poor ARI scores.
On the other hand, DBHT's topological constraints (bubble and converging bubble) 
uses global information, which makes them less sensitive to wrong agglomeration decisions. 
DBHT can also suffer from the sensitivity of agglomeration
decisions when the global information is not captured correctly (e.g., in data set 9).

\myparagraph{$k$-Means} 
\textsc{par-tdbht-1} and \textsc{par-tdbht-10} generate clusters of similar quality to \textsc{k-means}
across all data sets. 
However, \textsc{k-means} does not produce a dendrogram. %
To find clusters at different scales, we would have to
run the algorithm multiple times, which would result in longer running times.%

\begin{figure}[!t]
    \centering
    \vspace{-5pt}
    \includegraphics[width = 0.8\columnwidth]{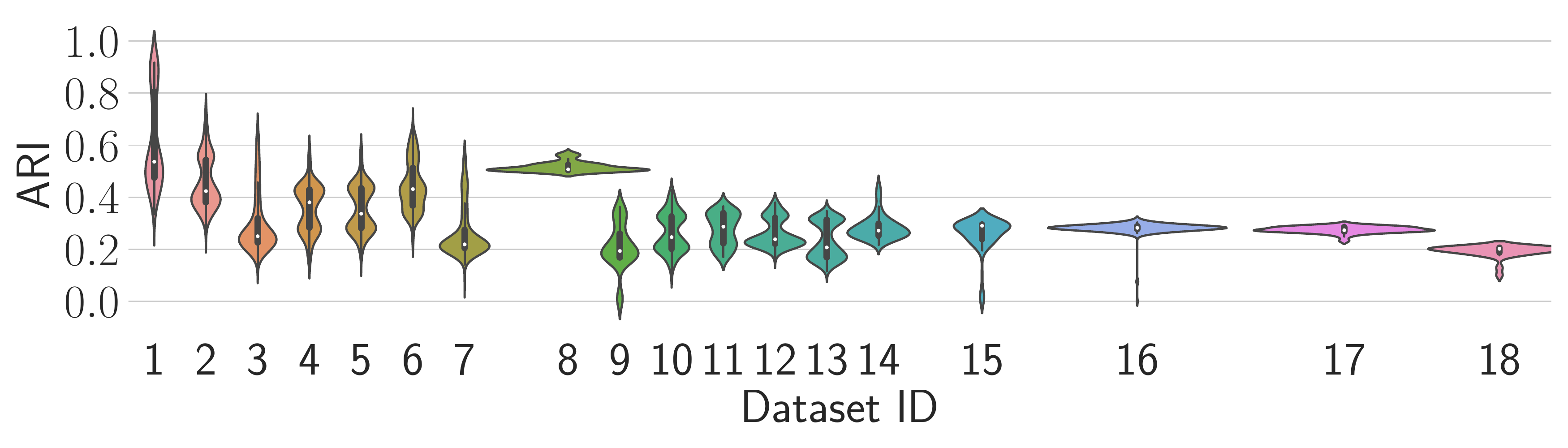}
    \includegraphics[width = 0.8\columnwidth]{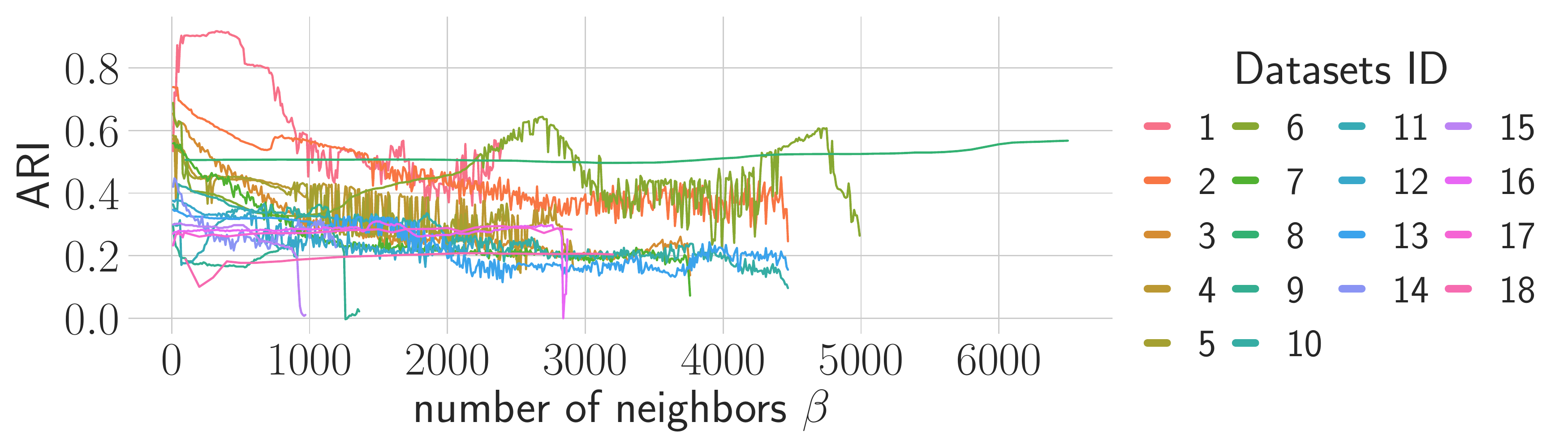}
    \caption{ARI of \textsc{k-means-s} with different number of nearest neighbors on each data set. The top plot shows the distribution of ARI scores for different values of $\beta$. The bottom plot shows the ARI scores vs.\ the value of $\beta$, demonstrating the oscillating behavior. }
    \label{fig:kmeans}
\end{figure}

Using the spectral embedding preprocessing step boosts the \textsc{k-means} method
to achieve the best quality on most data sets.
However, we show in
\cref{fig:kmeans} that the quality of \textsc{k-means-s}
is highly sensitive to the choice of parameter $\beta$ (the number of nearest neighbors) in many data sets.
On the top plot of \cref{fig:kmeans}, we see that for most data sets,
using different values of $\beta$ gives a wide range of ARI scores. On the bottom plot,
we see that the ARI oscillates with different $\beta$ for many data sets, and that the best $\beta$
is very different for each data set. 
As a result, this parameter is hard to choose apriori.
We also tried \textsc{par-tdbht} on the embedded data sets, and found the ARI scores to be similar to those of the non-embedded data sets. For data sets 8, 17, and 18, we run out of memory when running spectral embedding with $\beta$ values of 6600, 3000, and 4000 respectively. For all other data sets, we tested $\beta$ ranging from $10$ to $n$.

\begin{figure}[!t]
    \centering
    \vspace{-5pt}
    \includegraphics[width = 0.8\columnwidth]{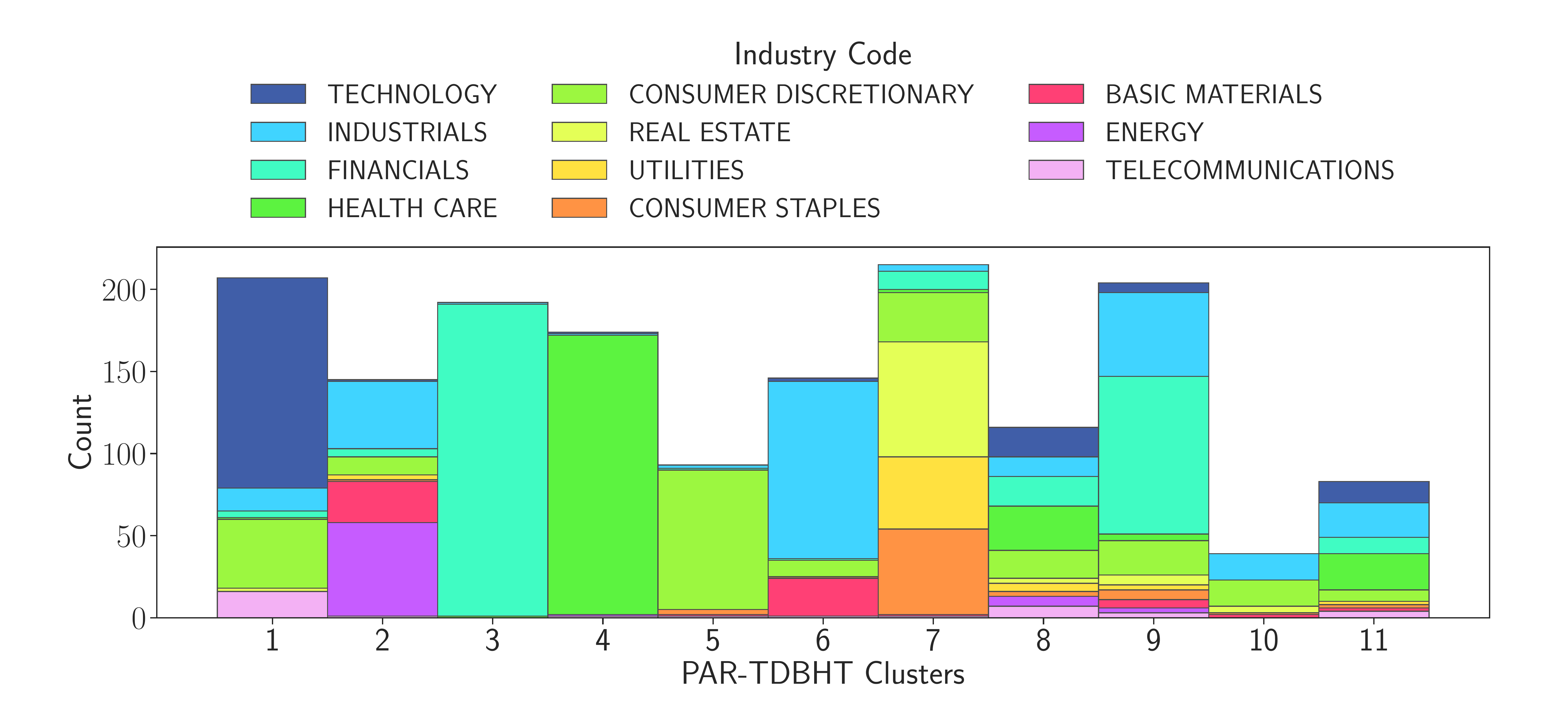}
    \caption{Clustering result on the US stock data set compared to ground truth using \textsc{par-tdbht} with a prefix of size 30.}
    \label{fig:stock}
\end{figure}

\myparagraph{Example: Clustering Stocks}
\cref{fig:stock} shows the clusters produced by \textsc{par-tdbht} with a prefix size of 30 on the US stock data set. 
We preprocess the daily stock prices by computing the detrended daily log-return using the method by Musmeci et al.~\cite{musmeci2015relation}. We then compute a spectral embedding of the normalized detrended daily log-returns.  
Finally, we compute the Pearson correlation of the embedded data and run \textsc{par-tdbht} on the correlation matrix.  
\revised{
We see that \textsc{par-tdbht} is able to find structure and patterns in the stock data. It accurately clusters the ``financial" stocks, ``health care" stocks, and ``consumer discretionary" stocks. There is also a cluster with mostly ``technology" stocks, and a cluster with mostly ``industrials" stocks. We also see that almost all of the ``energy", ``utilities", ``consumer staples", and ``real estate" stocks are clustered together.  Though for some clusters, the companies in the cluster do not come from a single industry in the ICB classification, they have commonalities between them.
For example, cluster 7 contains both consumer discretionary and consumer staple stocks---many companies in the consumer discretionary industry are restaurants and wholesale, and many companies in the consumer staple industry are food and drink companies, which are related. 
As another example, in cluster 1, many of the companies not in the Technology industry do most of their business on the internet. Most of the companies in cluster 1 from the Industrial industry are in the ``Electronic and Electrical Equipment" sector,\footnote{In the ICB, there are four levels of classification: industry, super-sector, sector, and sub-sector.} which is related to technology. 
}

\iffull
\showfull{

Clusters 8 and 9 are more mixed than other clusters because the companies in these clusters are smaller and so their stock prices might be more 
 volatile. We can see from Figure~\ref{fig:market-cap} that while there are no significant differences in the median market cap between the sectors  (Figure~\ref{fig:market-cap}(a)), the market cap of clusters 8 and 9 are in general lower than that of other clusters (Figure~\ref{fig:market-cap} (b)). }

\begin{figure}
    \centering

    \includegraphics[width = \columnwidth]{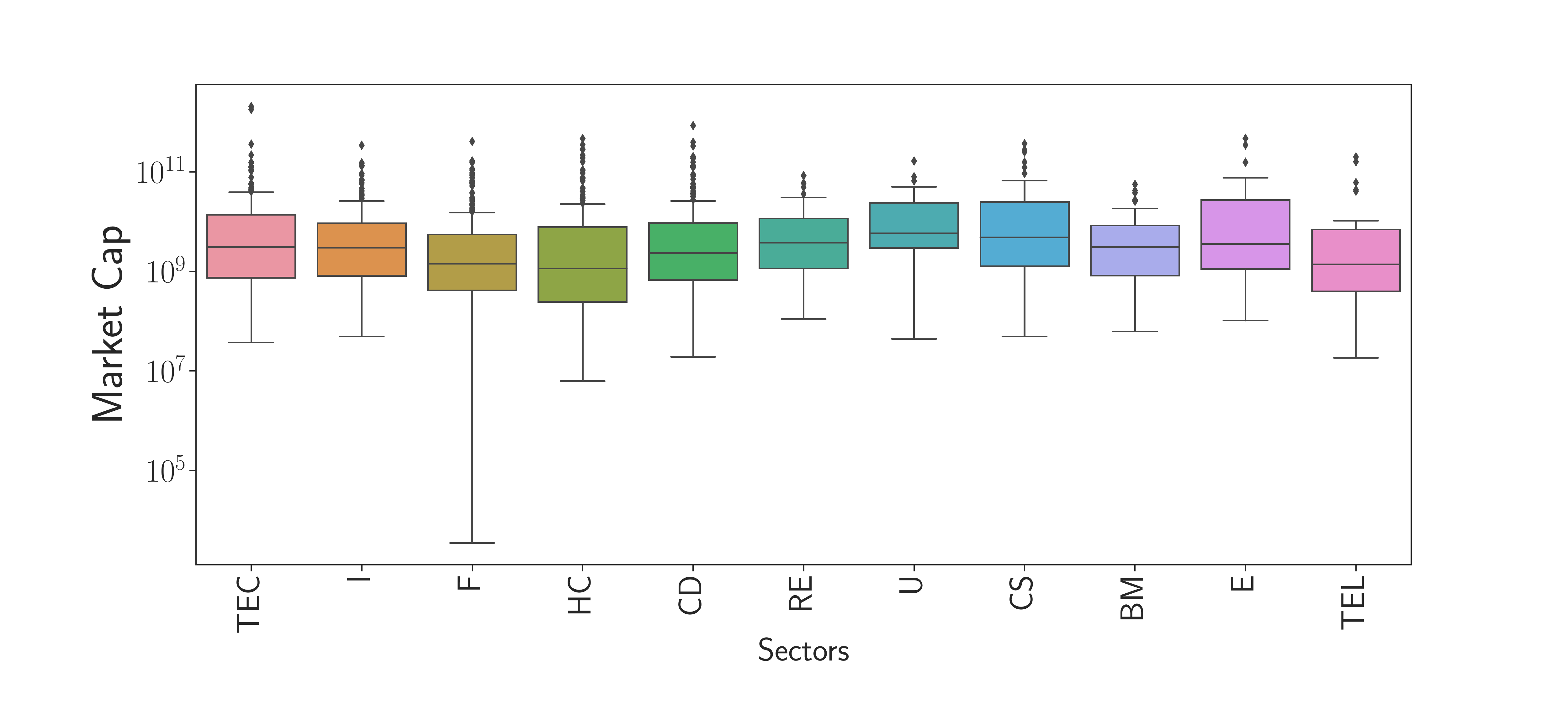}
(a)
    \includegraphics[width = \columnwidth]{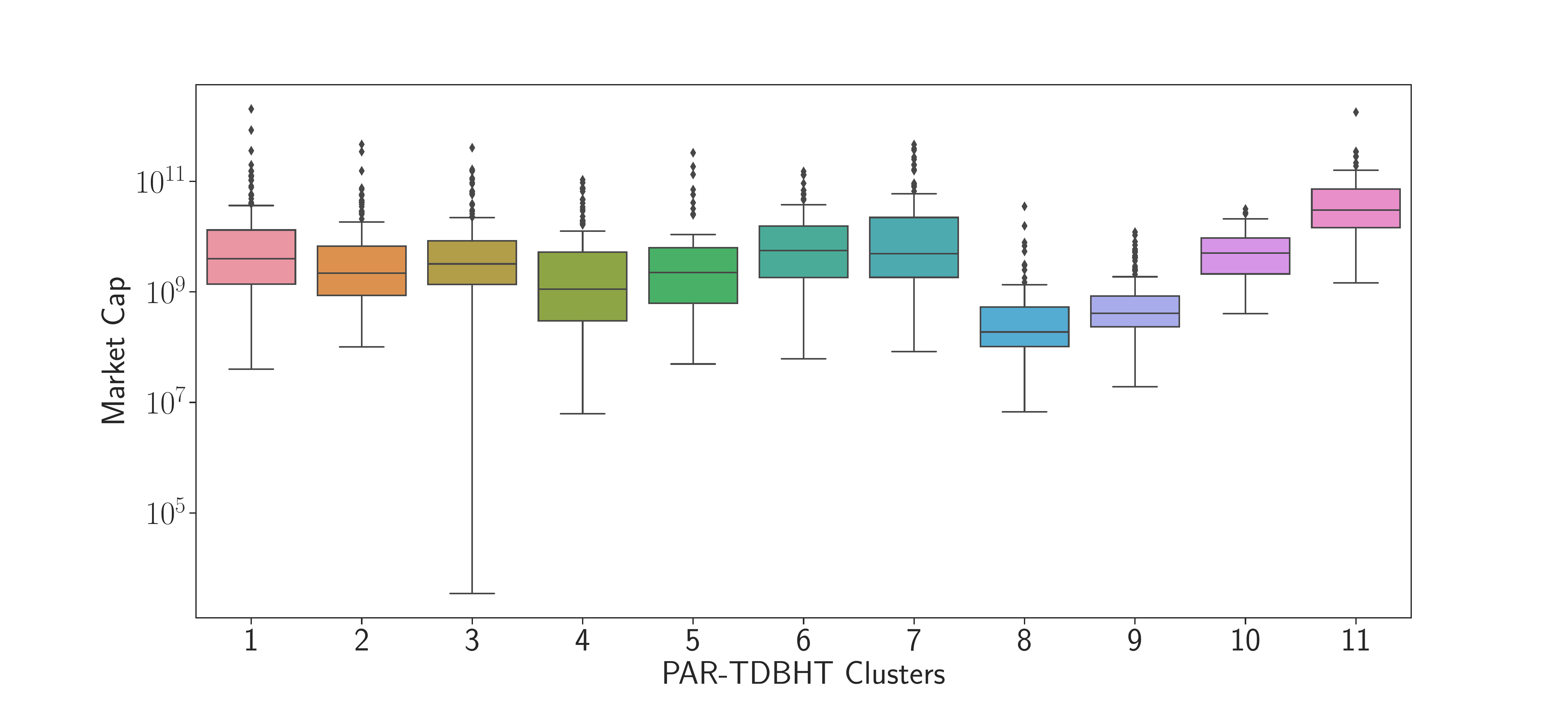}
(b)
    \caption{Market cap box plot of tickers in each sector (a) and each \textsc{PAR-TDBHT} cluster (b). The market cap information is obtained from Yahoo Finance.
    The sectors are marked with their initials. The corresponding sector names are in \Cref{tab:abbr}.
    }
    \label{fig:market-cap}
\end{figure}

\begin{table}
\centering
\caption{Abbreviation of sector names.}\label{tab:abbr}
\begin{tabular}{cc}
\toprule
Abbreviation & Name  \\ \hline 
TEC & Technology  \\
I & Industrials  \\
F & Financials  \\ 
HC & Health Care  \\ 
CD & Consumer Discretionary  \\
RE & Real Estate  \\
U & Utilities  \\
CS & Consumer Staples \\ 
BM & Basic Materials  \\
E & Energy  \\  
TEL & Telecommunications  \\
\bottomrule
\end{tabular}
\end{table}

\fi
The ARI score of this clustering is 0.36.
In comparison, the ARI score of the exact TMFG clustering is 0.28.

In practice, stock clustering can be used in portfolio optimization and risk hedging~\cite{musmeci2015relation, tola2008cluster}.

\myparagraph{Time and Quality Trade-off}
 \textsc{par-tdbht} provides a good trade-off between 
runtime and clustering quality. 
It is faster than the \textsc{k-means-s}
method, and is able to produce clusters of similar quality.
Although \textsc{par-tdbht} is slower than \textsc{avg} and \textsc{comp}, its clustering quality is more stable and in most cases
better. Furthermore,
it is able to finish within 30 seconds for all of the data sets.

%% file: conclusion.tex
\section{Conclusion}
We designed new parallel algorithms for
constructing TMFGs and DBHTs. 
We showed that our algorithms are usually able to produce better clusters than complete and average linkage clustering. 
We believe that our implementations will be a useful addition to the toolkit of data scientists who need to perform clustering efficiently and accurately.